\documentclass[pre,amsmath,amssymb,onecolumn, showpacs, superscriptaddress,10pt]{revtex4-2}

\usepackage[english]{babel}
\usepackage[utf8]{inputenc}
\usepackage[T1]{fontenc}
\usepackage{amsmath}
\usepackage[colorlinks=true,citecolor=blue]{hyperref}
\usepackage{graphicx}
\usepackage{amsfonts}
\usepackage{amsthm}

\newcommand{\be}{\begin{equation}}
\newcommand{\ee}{\end{equation}}
\newcommand{\bea}{\begin{eqnarray}}
\newcommand{\eea}{\end{eqnarray}}

\newcommand{\cP}{{\cal P}}

\usepackage{color}
\definecolor{Blue}{rgb}{0.00, 0.00, 1.00}
\definecolor{Red}{rgb}{1.00, 0.00, 0.00}
\definecolor{Green}{rgb}{0.00, 0.70, 0.00}

\begin{document}

\title{Burgers dynamics for Poisson point process initial conditions of the Weibull class}

\author{Patrick Valageas}
\affiliation{Universit\'e Paris-Saclay, CEA, CNRS, Institut de Physique th\'eorique, 91191 Gif-sur-Yvette, France}

\begin{abstract}

We derive the statistical properties of one-dimensional Burgers dynamics with
stochastic initial conditions for the velocity potential defined by a Poisson point process 
whose intensity follows a power law with exponent $\alpha > -1$.
This choice realizes a broad and analytically tractable family of strongly non-Gaussian
initial states, relevant as limiting cases for problems ranging from turbulence phenomenology
to simplified models of structure formation in cosmology.
Working in the inviscid limit and exploiting the geometrical construction of solutions in terms
of first-contact parabolas, we derive explicit analytical expressions for a broad set of statistical quantities.
These include the one- and two-point probability distributions of the velocity, the multiplicity functions 
of voids and shocks, and the velocity and density correlation functions together with their associated
power spectra.
We also show that the full hierarchy of $n$-point distributions factorizes into a sequence of 
two-point conditional probabilities.
This class of initial conditions leads to self-similar evolution and produces probability distributions
characterized by stretched-exponential tails, with tail exponents spanning the full range from unity 
to infinity. The associated characteristic length scale grows as a power law of time, with an exponent 
lying between zero and one half. 

\end{abstract}

\maketitle

\section{Introduction}

Nonlinear transport phenomena driven by advection and dissipation play a fundamental role in statistical 
physics. A classic model in this context is the Burgers equation \cite{Burgers1974,Hopf1950,Cole1951}, 
which provides a minimal framework where nonlinear steepening and dissipative effects both appear. 
In the inviscid regime—where the viscosity $\nu$ tends toward zero—velocity gradients grow increasingly 
sharp until shock discontinuities develop \cite{Burgers1974,Gurbatov1981,Whitham1999}, separated by 
linear velocity ramps. The resulting fields display strong intermittency and highly nonuniform density 
patterns. Consequently, Burgers turbulence serves as an important theoretical setting for studying 
nonlinear stochastic processes in far-from-equilibrium environments \cite{Frisch2001,Bec2007}, with 
applications ranging from turbulence phenomenology \cite{Kraichnan1968,Kida1979} to irreversible 
aggregation models \cite{Frachebourg2000,Valageas2009}. 
Despite its apparent simplicity, the Burgers equation exhibits a rich statistical 
structure when initialized with random conditions. A central difficulty is that nonlinear evolution 
generates singular structures (shocks) and highly non-Gaussian fluctuations, making it challenging to 
obtain explicit analytical results beyond a few special cases. Understanding how statistical properties 
depend on the choice of initial conditions and identifying classes where exact results can be derived
therefore remains an important open problem.

In this work, we investigate the deterministic Burgers equation without any external noise, so that randomness arises solely from stochastic initial conditions.
This deterministic inviscid Burgers equation also appears in cosmology, 
as the so-called "adhesion model" \cite{Gurbatov1989,Gurbatov1991,Vergassola1994,Valageas2011}
to model the formation of large-scale structures. 
The infinitesimal viscosity halts shell crossings and avoids the unphysical particle
escape from potential wells entailed by the simpler Zeldovich approximation 
\cite{Zeldovich1970,Shandarin1989}.
From a broader perspective, Burgers dynamics provide a tractable laboratory for studying how 
complex spatial structures—such as shocks and voids—emerge from stochastic initial conditions
for systems that are governed by ballistic processes. Insights gained in this context are relevant not 
only for turbulence and cosmology, but also for a wide range of nonlinear systems where clustering, 
intermittency, and scaling behavior play a key role, such as
exclusion-process-type models and related stochastic lattice gases \cite{Spohn-1991,Doyon-2025}.
Therefore, the present analysis could provide a natural testing ground for ideas developed in 
generalized hydrodynamics and suggest possible extensions where both ballistic and diffusive
effects come into play.

A significant portion of prior work considers Gaussian initial conditions with power-law spectra 
\cite{Gurbatov1981,She1992,Gurbatov1997}. These lead to self-similar evolutions with an
expanding integral scale. Although the Hopf–Cole transformation \cite{Hopf1950,Cole1951} gives an 
explicit expression for the solution at arbitrary time in terms of the initial state, closed-form 
statistical results are only available in special cases—for instance, for Brownian initial velocity 
fields \cite{Bertoin1998,Valageas2009a} or for white-noise initial velocities 
\cite{Frachebourg2000a,Valageas2009b}. 

Another class of initial conditions that yields explicit results is based on Poisson point processes. 
When the Poisson intensity follows a power law, the dynamics again become self-similar. 
Building on \cite{Molchanov1997,Gueudre2014}, we recently studied the case of negative exponents 
$\alpha < -3/2$ \cite{Valageas2025}, which leads to Fr\'echet-type probability distributions
with heavy power-law tails. 
In this paper, following \cite{Bernard1998,Bauer_1999}, we instead focus on exponents $\alpha > -1$, 
which produce Weibull-type distributions characterized by stretched-exponential tails. 
The motivation for focusing on this class is twofold. First, it complements the previously 
studied Fréchet regime and thereby completes the characterization of Poisson-driven initial conditions across all admissible exponents. Second, the Weibull class exhibits qualitatively different statistical behavior—most notably, rapidly decaying tails—which allows us to explore how the nature of rare events influences the resulting dynamics.
We extend previous analyses by computing one- and two-point statistics of both the velocity and density fields for arbitrary Poisson exponent $\alpha > -1$, and by providing numerical results for representative examples.

Our analysis emphasizes how the same geometric mechanism controls the distribution of 
velocities, the multiplicity of voids and shocks, and the behavior of correlation functions,
and clarifies which aspects depend on $\alpha$ and which are universal.
For the reader’s convenience, we now summarize the main results.
(i) We show that the dynamics remain exactly self-similar for all $\alpha>-1$,
and we obtain the explicit scaling of the integral length $L(t)$
and of the rescaled fields, thereby identifying the single growing length that controls the evolution.
(ii) We derive closed-form expressions for the one-point Eulerian velocity distribution and for the 
probability density of Lagrangian displacements and corresponding densities, and we show that their 
tails are governed by stretched exponentials whose exponents span the full range from 1 to $\infty$
as $\alpha$ is varied.
(iii) We compute the two-point Eulerian statistics of the velocity and density, including the 
associated correlation functions and power spectra, and we identify the stretched-exponential decay of 
correlations at large separations.
(iv) We obtain the void probability, the void multiplicity function, and the shock multiplicity 
function, which together characterize how matter is partitioned into empty regions and shocks, and we 
provide their asymptotic behavior for small and large intervals or masses.
(v) Finally, we show that the full hierarchy of $n$-point distributions factorizes into a sequence of 
two-point conditional probabilities, so that all higher-order statistics can be reconstructed from a 
finite set of building blocks.

This paper is organized as follows.
In Section~\ref{sec:Initial}, we review the equations of motion, the geometric interpretation of the system, and the class of initial conditions considered, and we show two numerical realizations illustrating the dependence on the exponent $\alpha$.
Section~\ref{sec:one-point-Eulerian} derives the one-point Eulerian velocity distribution $P_0(v)$.
In Section~\ref{sec:two-point}, we present the two-point Eulerian distributions of the velocity and density fields, along with the void distribution and the energy power spectrum.
Higher-order distributions are briefly discussed in Section~\ref{sec:higher-order}.
In Section~\ref{sec:Lagrangian}, we turn to the Lagrangian statistics of particle displacements and derive the shock multiplicity function.
We conclude in Section~\ref{sec:conclusion}.

\section{Equations of motion and Initial conditions}
\label{sec:Initial}

In this section we recall the standard Burgers framework and fix our notation, 
with the aim of making the probabilistic constructions easier to follow for readers
not already familiar with the Hopf–Cole and geometric formulations.
Our strategy is to rewrite the dynamics into a simple variational problem, which will directly
connect to the Poisson point process initial conditions.

\subsection{Equations of motion}
\label{sec:Equations-of-motion}

We consider in this article the one-dimensional Burgers equation \cite{Burgers1974} for the velocity 
field $v(x,t)$ in the limit of vanishing viscosity,
\be
\frac{\partial v}{\partial t} + v \frac{\partial v}{\partial x} = \nu \frac{\partial^2 v}{\partial x^2}
\hspace{1cm} \mbox{with} \hspace{1cm} \nu \rightarrow 0^+ ,
\label{eq:Burg}
\ee
together with the associated density field $\rho(x,t)$, which satisfies the continuity equation
for an initially uniform density $\rho_0$ \cite{Gurbatov2011}.
As is well known \cite[]{Hopf1950,Cole1951,Gurbatov1991}, introducing the velocity potential 
$\psi(x,t)$ defined by $v= \partial\psi/\partial x$, and applying the transformation
$\psi(x,t)= - 2\nu\ln\theta(x,t)$, reduces the Burgers equation to the linear heat equation for 
$\theta$. This yields the explicit solution 
\be
v(x,t) = \frac{\partial\psi}{\partial x} \hspace{0.5cm} \mbox{with} \hspace{0.5cm} 
\psi(x,t)= - 2\nu \ln \int_{-\infty}^{\infty} \frac{d q}{\sqrt{4\pi\nu t}} \;
\exp\left[-\frac{(x-q)^2}{4\nu t} - \frac{\psi_0(q)}{2\nu}\right] ,
\label{eq:psinu}
\ee
where $\psi_0(q)=\psi(q,t=0)$ is the initial potential. 
In the limit $\nu \rightarrow 0^+$, the steepest-descent method gives \cite{Gurbatov1991,Bec2007}
\be
\psi(x,t) = \min_q \left[ \psi_0(q) + \frac{(x-q)^2}{2t} \right]
\hspace{0.5cm} \mbox{and} \hspace{0.5cm} v(x,t) = \frac{x-q(x,t)}{t} ,
\label{eq:psinu0}
\ee
where the Lagrangian coordinate $q(x,t)$ denotes the point at which the minimum is achieved.
Eulerian positions $x$ for which the minimization problem admits two solutions, $q_-<q_+$,
correspond to shock locations, and all mass initially between $q_-$ and $q_+$ collapses to the point
$x$. The mapping $q \mapsto x(q,t)$ is known as the Lagrangian map, whereas its inverse 
$x \mapsto q(x,t)$ is the inverse Lagrangian map.
They are discontinuous at the locations of voids and shocks \cite{Bec2007,Vergassola1994}.

\subsection{Geometrical interpretation and Legendre transform}
\label{sec:Geometrical-interpretation}

As is well known \cite{Burgers1974,Gurbatov1991}, the minimization problem in Eq.(\ref{eq:psinu0})
has a simple geometrical interpretation. Consider the downward-opening parabola $\cP_{x,c}(q)$
centered at $x$ and of maximum value $c$, defined by
\be
\cP_{x,c}(q) = c - \frac{(q-x)^2}{2 t} .
\label{eq:paraboladef}
\ee
Starting from a very negative value of $c$, so that the parabola lies entirely below $\psi_0(q)$, 
we increase $c$ until the parabola first touches the initial potential.
The abscissa of this point of first contact gives the Lagrangian coordinate
$q(x,t)$ and the corresponding potential is $\psi(x,t)=c$.
This geometric viewpoint is central to the rest of the paper: instead of solving a partial
differential equation, we will characterize the statistics of the first-contact points
between parabolas and the initial potential.
This will determine all subsequent Eulerian and Lagrangian statistics.

The expression (\ref{eq:psinu0}) for the velocity potential can also be written
using a Legendre-Fenchel transform \cite{She1992,Vergassola1994,Valageas2011}.
Define the linear Lagrangian potential $\varphi_L(q,t)$ and the associated Lagrangian map
$x_L(q,t)$ by
\be
\varphi_L(q,t) = \frac{q^2}{2} + t \psi_0(q), \;\;\;
x_L(q,t) = \frac{\partial\varphi_L}{\partial q} = q + t v_0(q) ,
\ee
which describe the evolution in the absence of shocks.
Introducing
\be
H(x,t) = \frac{x^2}{2} - t \psi(x,t) ,
\ee
the minimization in Eq.(\ref{eq:psinu0}) becomes
\be
H(x,t) = \max_q \left[ x q - \varphi_L(q,t) \right] = {\cal L}_x[ \varphi_L(q,t) ] ,
\label{eq:H-def}
\ee
where ${\cal L}_x$ denotes the Legendre transform evaluated at $x$.
In this way, $\psi(x,t)$ is obtained from $\psi_0(q)$ through a Legendre transform,
which simultaneously yields the inverse Lagrangian map $q(x,t)$ and the velocity field $v(x,t)$.

\subsection{Initial condition}
\label{sec:Initial-condition}

In this paper, following \cite{Bernard1998}, 
we consider stochastic initial conditions for $\psi_0(q)$ generated by a Poisson
point process with intensity $\lambda(\psi_0)$ in the upper half-plane $(q,\psi_0>0)$.
Thus, the initial state consists of a set of points $\{(q,\psi_0)_i\}$, and the probability 
of finding $n$ points within any domain $\cal B$ is given by the Poisson law
\be
P(N_{\cal B}=n) = \frac{\Lambda({\cal B})^n}{n!} e^{-\Lambda({\cal B})} , \;\;\;
\mbox{with} \;\;\; \Lambda({\cal B}) = \int_{\cal B} dq d\psi_0 \, \lambda(\psi_0) .
\ee
As in \cite{Bernard1998}, we focus on the case where the intensity $\lambda(\psi_0)$ follows
a power-law,
\be
\psi_0 > 0 : \;\;\; \lambda(\psi_0) = a \, \psi_0^{\alpha} , \;\;\; a > 0 , \;\;\; \alpha > -1 , \;\;\;
\mbox{and for}  \;\;\;  \psi_0 \leq 0 : \;\;\; \lambda(\psi_0) = 0 .
\label{eq:lambda-psi}
\ee
The condition $\alpha>-1$ ensures that $\Lambda({\cal B}_{x,c})$ is finite when ${\cal B}_{x,c}$
denotes the region below the parabola $\cP_{x,c}(q)$ defined in Eq.(\ref{eq:paraboladef}), with $c>0$.
In what follows, we study the velocity and density fields obtained from Eq.(\ref{eq:psinu0})
for the power-law intensity (\ref{eq:lambda-psi}).
Because $\lambda(\psi_0)$ is independent of $q$, the system remains statistically
homogeneous at all times.
The choice of a Poisson point process in the $(q,\psi_0)$-plane is motivated by its simplicity
and flexibility: it generates strongly non-Gaussian initial conditions with tunable tail behavior, 
while remaining amenable to exact calculations.
The power-law form of $\lambda(\psi_0)$ is what ultimately leads to Weibull-type distributions
for the evolved fields when $\alpha > -1$.

This Poisson point process gives a discrete set of points $\{(q,\psi_0)_i\}$.
Although this does not define a continuous initial velocity potential $\psi_0(q)$, one may formally
construct such a function by drawing, from each point $\{(q,\psi_0)_i\}$, two nearly vertical line segments
with slopes $\pm \gamma$ extending upward to infinity, and then connecting the resulting narrow
triangular regions.
For a finite number of points within a bounded interval $[q_1,q_2]$, and upon restricting to points with 
$\psi_{0,i} \leq \psi_{\max}$, this procedure yields a continuous function $\psi_0(q)$. 
Then, in the limits $\gamma \to \infty$ and $\psi_{\max} \to \infty$, these auxiliary vertical segments 
become irrelevant, as the first-contact parabolas defined in Eq.(\ref{eq:paraboladef}) intersect only the
original lower summits $\{(q,\psi_0)_i\}$.
However, thanks to the geometrical construction in Eqs.(\ref{eq:psinu0})-(\ref{eq:paraboladef}),
this intermediate regularization is unnecessary.
The minimization in Eq.(\ref{eq:psinu0}) can be taken directly over the discrete set of points
$\{(q,\psi_0)_i\}$, which fully determines the evolved potential $\psi(x,t)$ for all times $t>0$.
One could as well define the initial condition by this potential computed at a time $t_i>0$.
Thus, the detailed regularization of $\psi_0(q)$ 
is irrelevant for the dynamics: only the discrete set of Poisson points matters, because the
first-contact parabolas always touch the lower envelope of the configuration.
This ensures that the probabilistic description can remain purely discrete, which greatly simplifies 
the derivation of all subsequent statistical quantities.

The case of negative exponents $\alpha < -3/2$, which leads to Fr\'echet-type probability distributions,
was studied in \cite{Valageas2025}. It is associated with a change of sign for the definition of the
potential $\psi$, so that the analysis deals with first-contact points of the Poisson process with
upward parabolas, whereas here we encounter first-contact points with downward parabolic arcs.
This is because for $\alpha>-1$ the number of points grows (or decays very slowly) at large
distance over the horizontal axis, for $\psi_0 \gg 1$, so that first-contact points only make sense
with respect to downward parabolas that move upward from $c=-\infty$ (any upward parabola
at finite distance would have already crossed an infinite number of Poisson process points).
Instead, for $\alpha < -3/2$ the number of points quickly decreases at large distance but 
grows very fast close to the horizontal axis. Then, it only makes sense to consider first-contact 
points with respect to upward parabolas that move downward from $c=\infty$ (whereas downward
parabolas would always have their first-contact point at $c=0$).
Cases $-3/2 \leq \alpha \leq -1$ are excluded because they are not associated with a well-defined
system. The Poisson distribution is too broad, both with respect to $\psi_0 \to \infty$ and
$\psi_0 \to 0$, to permit a meaningful construction in terms of first-contact points with either
upward or downward parabolas.
This can also be seen from the scaling of the integral length scale in Eq.(\ref{eq:L-t}) below.
For $-3/2 \leq \alpha \leq -1$ it would decrease with time but become infinite for $t \to 0$.
This means that for such values of $\alpha$ it is not possible to define
from the discrete Poisson point process the dynamics of a continuous velocity field
at times $t>0$.

\subsection{Self-similar dynamics}
\label{sec:self-similar}

We introduce the rescaled coordinates
\be
x = a^{-1/(2\alpha+3)} t^{(\alpha+1)/(2 \alpha+3)} X , \;\;\; 
v= a^{-1/(2\alpha+3)} t^{-(\alpha+2)/(2\alpha+3)} V , \;\;\;
\psi = a^{-2/(2\alpha+3)} t^{-1/(2\alpha+3)} \Psi .
\label{eq:re-scaling}
\ee
The initial potential $\psi_0$ is rescaled in the same way to $\Psi_0$, for any given time $t$.
This yields, at all times,
\be
\Psi(X,t) = \min_{i} \left[ \Psi_{0,i} + \frac{(X-Q_i)^2}{2} \right]
\hspace{0.5cm} \mbox{and} \hspace{0.5cm} V(X,t) = X-Q_{i\star} ,
\label{eq:psi-Qi}
\ee
where $i_\star$ denotes the Poisson point for which the minimum is attained.
The intensity measure of the Poisson process becomes
\be
\Lambda({\cal B}) = \int_{\cal B} dQ d\Psi_0 \, \lambda(\Psi_0) \;\;\; \mbox{with} \;\;
\lambda(\Psi_0) = \Psi_0^{\alpha} \; \mbox{for} \; \Psi_0 > 0 , \;\; 
\lambda(\Psi_0) = 0 \; \mbox{for} \; \Psi_0 \leq 0 .
\label{eq:Lambda-Psi0}
\ee
Because this rescaling removes all explicit dependence on time, the dynamics are statistically
self-similar for all $t>0$.
In particular, the integral length scale $L(t)$—defined, for instance, as the scale separating 
the large- and small-scale asymptotic regimes—grows with time as
\be
L(t) \propto t^{\gamma} \;\; \mbox{with} \;\; \gamma = \frac{\alpha+1}{2\alpha+3} , \;\;\;
0 < \gamma < \frac{1}{2} .
\label{eq:L-t}
\ee
Thus, $L(t)$ always grows more slowly than $\sqrt{t}$, the limit attained as $\alpha \to \infty$.
The normalization constant $a$ is also absorbed by the rescaling (\ref{eq:re-scaling}),
so the dynamics depend only on the exponent $\alpha$.
Thus, the exponent $\gamma$ quantifies how fast structures merge and grow as the system evolves: 
smaller values of $\gamma$ correspond to a slower coarsening of shocks and voids, whereas the limit
$\gamma \to 1/2$ $(\alpha \to \infty)$ approaches the diffusive scaling of standard random-walk 
problems.

In the following, we focus on equal-time statistics and therefore work exclusively with the rescaled variables (\ref{eq:re-scaling}), using lowercase letters for simplicity.

\subsection{Numerical realizations}
\label{sec:realizations}

\begin{figure}
\centering
\includegraphics[height=5.cm,width=0.32\textwidth]{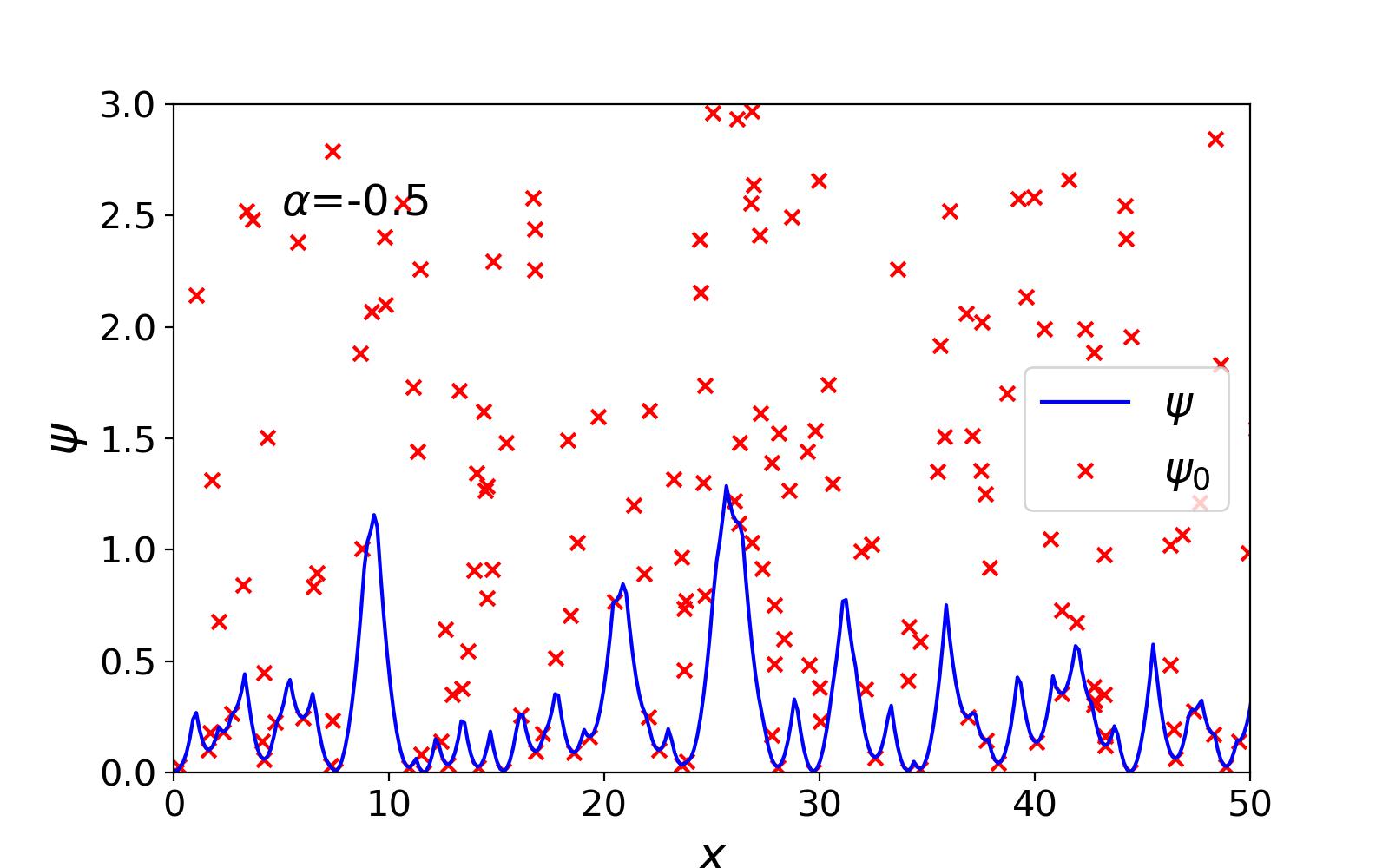}
\includegraphics[height=5.cm,width=0.32\textwidth]{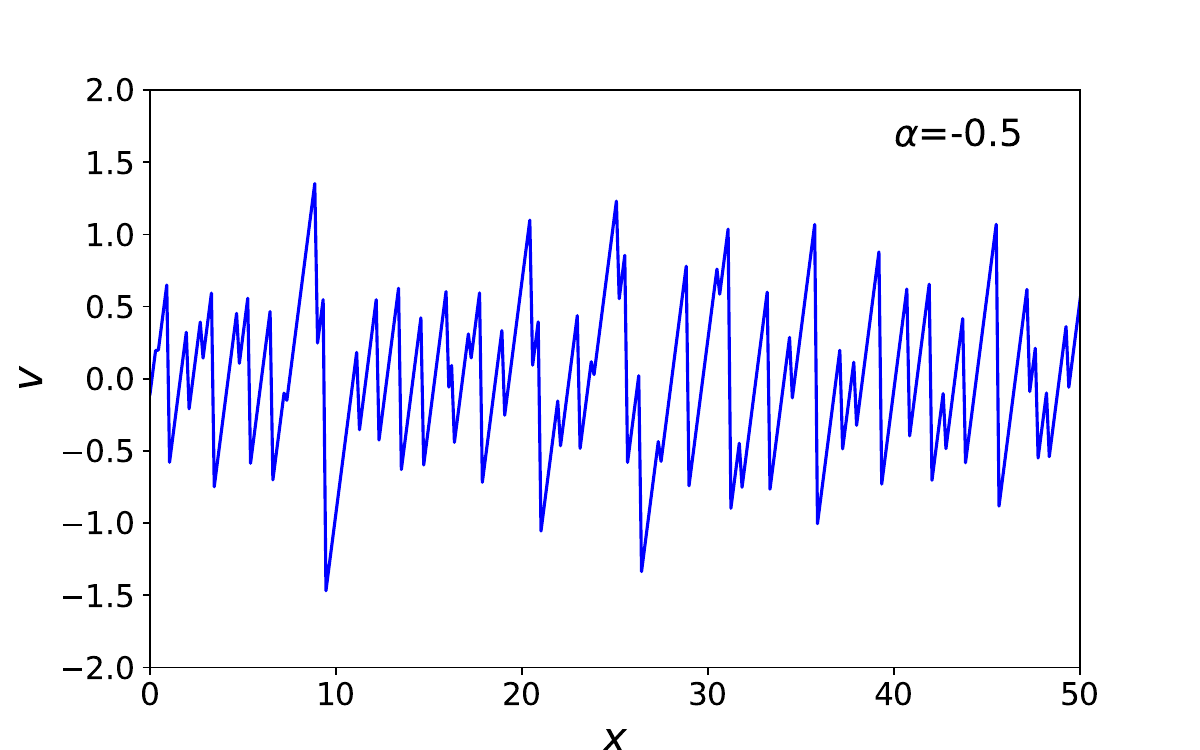}
\includegraphics[height=5.cm,width=0.32\textwidth]{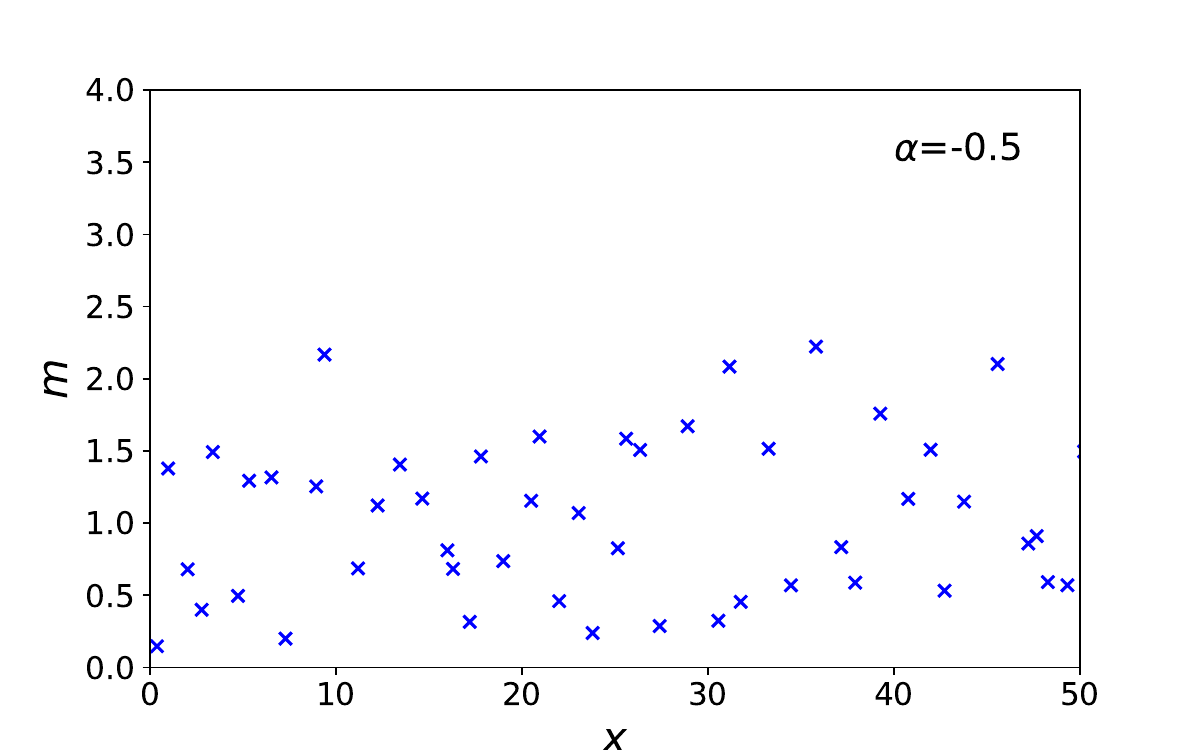}
\\
\includegraphics[height=5.cm,width=0.32\textwidth]{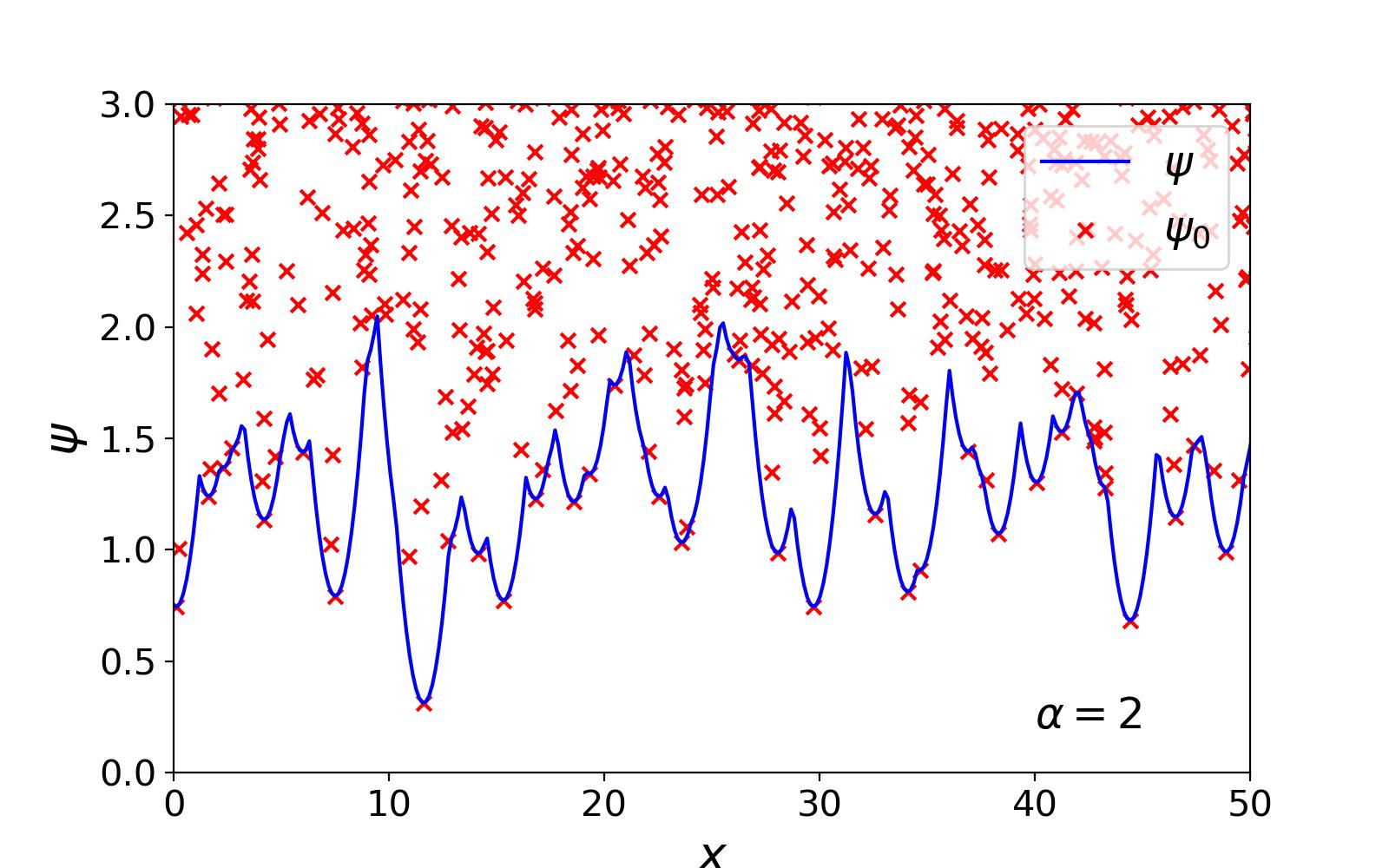}
\includegraphics[height=5.cm,width=0.32\textwidth]{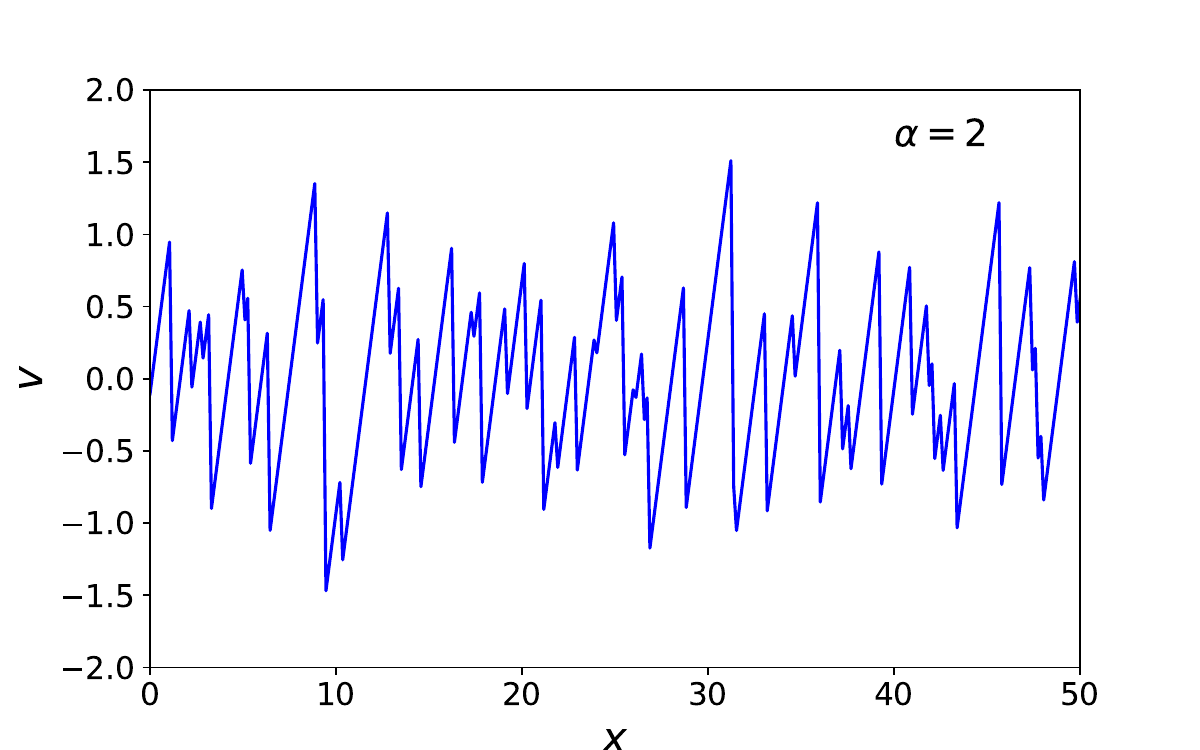}
\includegraphics[height=5.cm,width=0.32\textwidth]{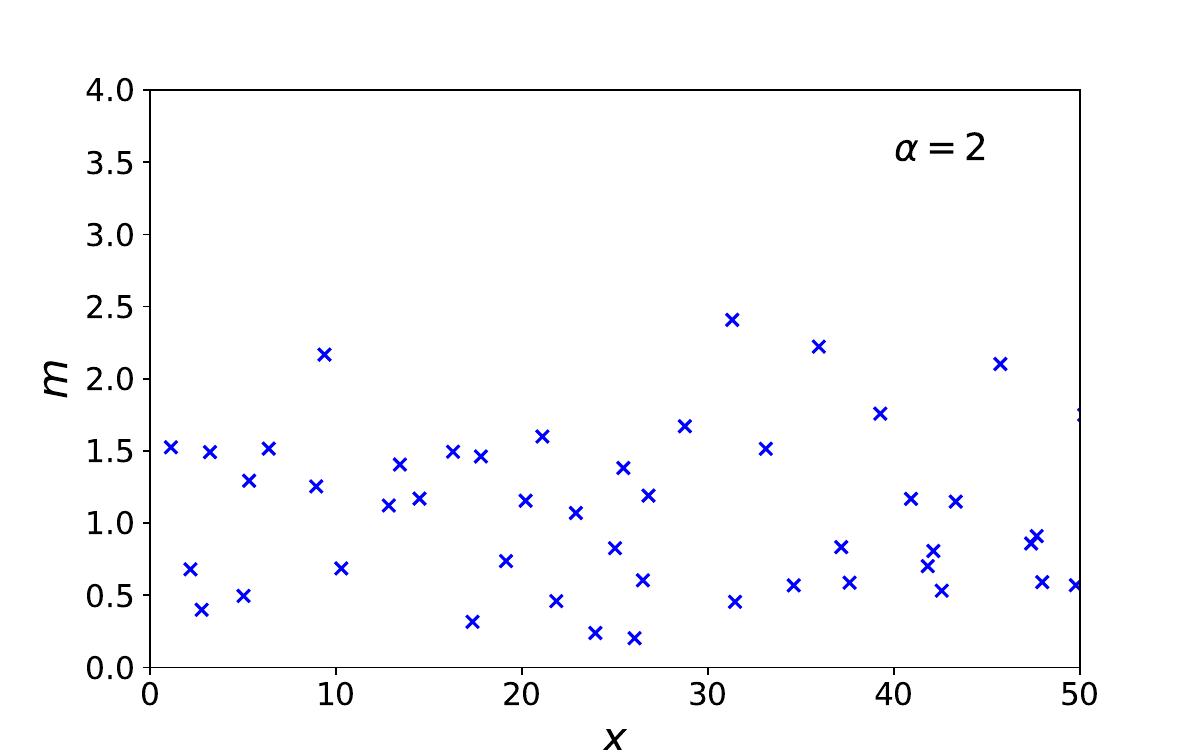}
\caption{A realization of the system for the cases $\alpha=-1/2$ (top row) and $\alpha=2$
(bottom row) at time $t=1$.
{\it Left column:} the initial velocity potential $\psi_0(x)$ (red crosses) and the evolved 
velocity potential $\psi(x,t)$ (blue solid curve) at time $t=1$.
{\it Middle column:} velocity field $v(x,t)$.
{\it Right column:} mass and location of the shocks.
}
\label{fig:realization}
\end{figure}

Figure~\ref{fig:realization} shows numerical realizations of the system for two representative
cases, $\alpha=-0.5$ (top row) and $\alpha=2$ (bottom row).
We use the rescaled coordinates (\ref{eq:re-scaling}), which also correspond to the choice $t=1$
and $a=1$.
The initial potential $\psi_0(q)$ is generated directly from a random number routine, using the fact 
that the Poisson point process (\ref{eq:lambda-psi}) becomes homogeneous in the $(q,y)$ plane
under the change of variable $y=\psi_0^{\alpha+1}/(\alpha+1)$. 
The potential $\psi(x,t)$ and velocity $v(x,t)$ at time $t$
are then obtained through the Legendre transform (\ref{eq:H-def}).
These examples are included to give an intuitive picture of the geometrical construction and of 
the typical configurations generated by different values of $\alpha$, before turning to formal 
probability distributions.
They also illustrate how the same qualitative pattern of ramps and shocks emerges in all cases,
while the distribution of amplitudes and void sizes is controlled by the Poisson exponent.

The initial velocity potential $\psi_0$ is highly singular, being defined by a Poisson point process
that provides the initial points $\{(q,\psi_0)_i\}$ shown by the red crosses.
The evolved potential $\psi(x,t)$ consists of a family of upward parabolic arcs, as predicted
by Eq.(\ref{eq:psinu0}), with their minima located at the deepest minima of $\psi_0$.
We can check that there are no initial points $(q,\psi_0)_i$ left below the evolved potential 
$\psi(x,t)$.
For the larger value of $\alpha$ the potential $\psi$ becomes more narrowly concentrated around
$\psi \simeq 1$.

The velocity field $v(x,t)$ displays the characteristic linear ramps $x/t$ of Burgers dynamics, 
separated by shocks corresponding to downward jumps. 
The visual appearance of the velocity field, as well as the shock positions $x_s$ and their masses 
$m_s=\Delta q$ (marked by the crosses in the right panels), is similar for both values of $\alpha$.
This is because the key statistical properties of the system—such as the void distribution
and the shock multiplicity function—exhibit comparable behaviors and differ mainly in the exponent
of their stretched-exponential tails.
These rare-event tails are difficult to discern by eye, and the most apparent feature in
Fig.~\ref{fig:realization}, common to all $\alpha$, is the steep decay of the probability
distributions, which confines most fluctuations to a finite range.
This stands in contrast with the Fr\'echet-type initial conditions studied in \cite{Valageas2025},
which produce heavy power-law tails that are easily visible and distinguishable in such plots.

In the following Sections, we derive analytical expressions for the probability distributions
of displacement, velocity, and density fields, as well as the statistics of voids and shocks
associated with these dynamics.

\subsection{Limit $\alpha \to \infty$}
\label{sec:alpha-infty}

In the limit $\alpha \to \infty$, the Poisson intensity (\ref{eq:Lambda-Psi0}) implies that
very few points lie below $\Psi_0=1$, while many lie above.
Consequently, the first-contact parabolas have $c \simeq 1$,
a trend already visible in the realizations for $\alpha=-0.5$ and $2$ shown in
Fig.~\ref{fig:realization}.
On the other hand, from Eq.(\ref{eq:Nvoid-asymp}) below, the number of voids per unit length
grows as $\sqrt{\alpha}$, indicating that the typical length scale decreases as $1/\sqrt{\alpha}$.
Therefore, in this limit, it is natural to rescale lengths, velocities and potentials as
\be
q = \tilde q / \sqrt{\alpha} , \;\;\; x = \tilde x / \sqrt{\alpha} , \;\;\; 
v = \tilde v / \sqrt{\alpha} , \;\;\; \psi = 1 + \tilde\psi/\alpha .
\label{eq:tilde-def}
\ee
One can explicitly check that by substituting these rescaled variables into the results derived
in the following Sections, one indeed obtains finite probability distributions.
Moreover, these expressions coincide with the $\alpha\to\infty$ limit obtained in
\cite{Valageas2025} for Fr\'echet-class initial conditions (corresponding to 
$\alpha\to -\infty$ in the notation of Eq.(\ref{eq:lambda-psi})).
In addition, they also match the late-time results obtained for Gaussian initial conditions with 
vanishing large-scale power \cite{Kida1979,Gurbatov1981,Gurbatov1991},
$E_0(k) \propto k^n$ with $n>1$ \cite{Gurbatov1997},
or in the hyperbolic asymptotic scaling \cite{Molchanov1995}.
By considering the Poisson point process (\ref{eq:lambda-psi}) for the initial potential itself,
we generalize this regime to the full class $\alpha > -1$, producing stretched-exponential tails
that vary with $\alpha$.
Furthermore, it is no longer necessary to focus on a late-time asymptotic regime, since
the dynamics are now fully self-similar.
However, whereas the stretched-exponential tails such as (\ref{eq:P0-large-q}) have an exponent
that diverges for $\alpha\to\infty$, the rescaling (\ref{eq:tilde-def}) yields Gaussian tails
in the limit $\alpha\to\infty$.
The explicit expressions obtained in this limit can be found in \cite{Valageas2025}, so
we do not consider it further in this article.

\section{One-point Eulerian distributions}
\label{sec:one-point-Eulerian}

\begin{figure}
\centering
\includegraphics[height=6cm,width=0.48\textwidth]{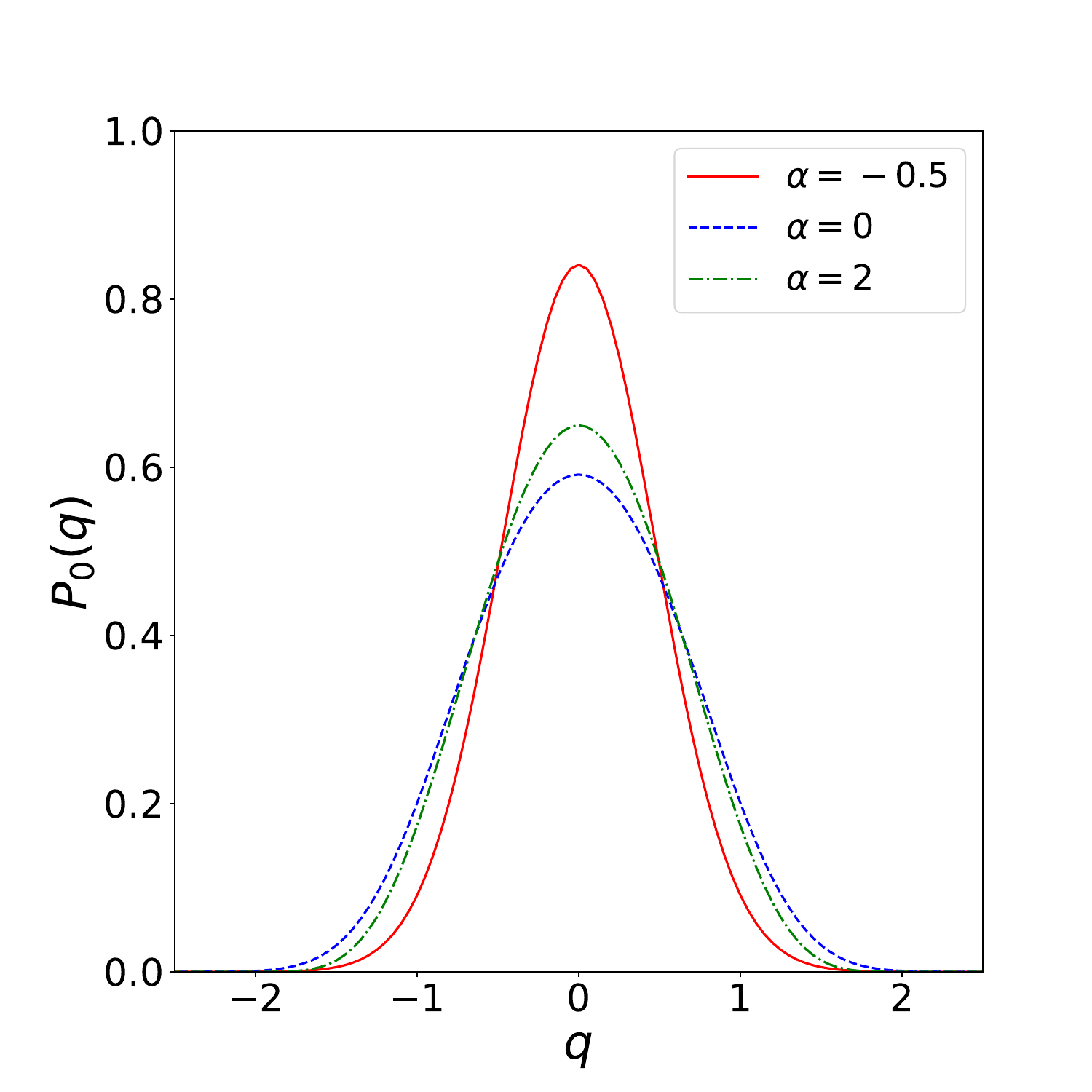}
\includegraphics[height=6cm,width=0.48\textwidth]{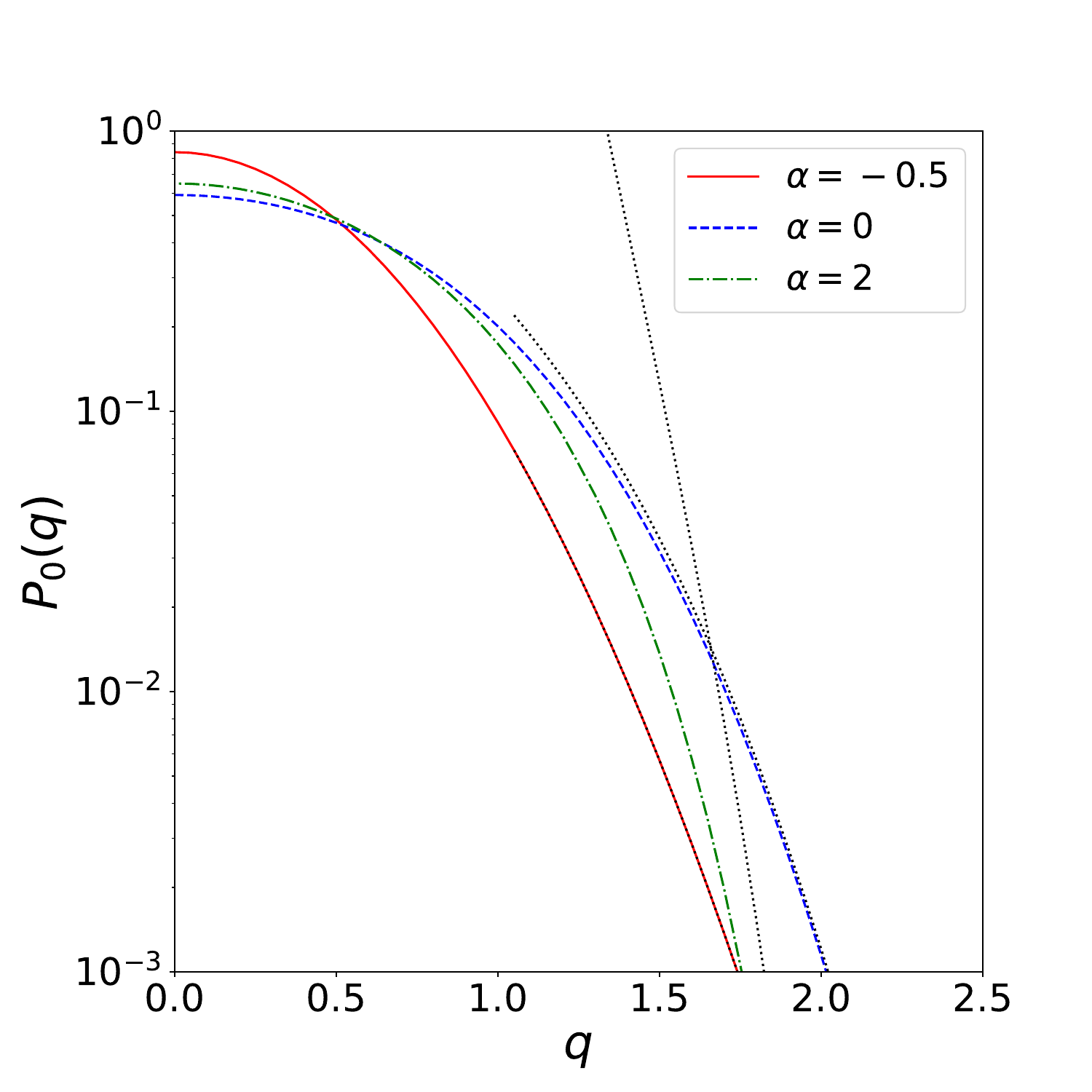}
\caption{
One-point probability distribution $P_0(q)=P_0(v)$ of the Lagrangian coordinate
$q$, or of the velocity $v$, from Eq.(\ref{eq:P_0-q}). 
We display our results on a linear scale (left panel) and a logarithmic scale (right panel), for the
cases $\alpha= -0.5, 0$, and $2$. 
The dotted lines in the right panel are the asymptotic results (\ref{eq:P0-large-q}).
}
\label{fig:P0_q}
\end{figure}

We begin with one-point statistics in this Section, because they already capture the basic
imprints of the Poisson–Weibull initial conditions on the evolved velocity and density fields,
such as the typical scale of velocity fluctuations and the nature of rare events.
They also serve as a building blocks for higher-order statistics.
Thus, we consider the one-point probability distribution $P_x(q)$ of the Lagrangian
coordinate $q$ corresponding to a given Eulerian position $x$. This also gives the distribution
of the velocity, $P_x(v)$, with $v=x-q$.
Owing to the statistical translational invariance, it is sufficient to focus on $x=0$,
since $P_x(q)$ depends only on $|q-x|$, so that $P_x(q) = P_0(|q-x|)$.
Then, the probability $P_0(q,c) dq dc$ that the first-contact point occurs at Lagrangian
coordinate $q$ with parabola height $c$ reads
\be
P_0(q,c) = \lambda(c-q^2/2) \, e^{- \Lambda_{\alpha} c^{\alpha+3/2} } \;\;\; \mbox{with} \;\;\;
\Lambda_{\alpha} = \frac{\sqrt{2\pi} \Gamma(\alpha+1)}{\Gamma(\alpha+5/2)} .
\ee
This gives for the probability distribution $P_0(q)$
\be
P_0(q) = \int_{q^2/2}^{\infty} dc \, (c-q^2/2)^{\alpha} \, e^{- \Lambda_\alpha c^{\alpha+3/2}} .
\label{eq:P_0-q}
\ee
This distribution is even in $q$, it has a finite value at the origin,
\be
P_0(q=0) = \frac{1}{\alpha+1} \Lambda_{\alpha}^{-(2 \alpha+2)/(2\alpha+3)} \, 
\Gamma \left( \frac{4\alpha+5}{2\alpha+3} \right)  ,
\label{eq:P0_q=0}
\ee
and a stretched-exponential tail at large distance
\be
q \gg 1 : \;\;\; P_0(q) \simeq \Gamma(\alpha+1) \left( (2\alpha+3) \Lambda_{\alpha} 2^{-\alpha-3/2}
q^{2\alpha+1} \right)^{-\alpha-1} e^{-\Lambda_{\alpha} 2^{-\alpha-3/2} q^{2\alpha+3} } .
\label{eq:P0-large-q}
\ee
This makes transparent how changing $\alpha$ reshapes the probability of large displacement
or velocity fluctuations through modified stretched-exponential tails.
Since for $x=0$ we have $v=-q$, the velocity probability distribution $P_0(v)=P_0(q=-v)$ is given by the
same expression.
In the cases $\alpha=-1/2$ and $0$ the integral (\ref{eq:P_0-q}) reads
\be
\alpha= -1/2 : \;\;\; L(t) \propto t^{1/4} , \;\;\;
P_0(q) = 2^{-1/4} e^{-\pi q^2 / \sqrt{2} }  ,
\ee
\be
\alpha= 0 : \;\;\; L(t) \propto t^{1/3} , \;\;\;
P_0(q) =  2^{-2/3} 3^{-1/3} \, \Gamma\left( \frac{2}{3}, \frac{2 q^3}{3} \right) .
\ee

Figure~\ref{fig:P0_q} shows the curves $P_0(q)$ for the cases $\alpha=-0.5, 0$, and $2$.
These same values of $\alpha$ will also be used in the subsequent figures throughout this article.
We verify that our numerical evaluation of Eq.(\ref{eq:P_0-q}) is consistent with the asymptotic
stretched-exponential tail given in (\ref{eq:P0-large-q}).
As already observed in the sample realizations shown in Fig.~\ref{fig:realization},
the velocity distribution is qualitatively similar for different values of $\alpha$.
It is mainly characterized by a bell-shaped peak at $v=0$, followed by a monotonic decrease for 
$|v|>0$ and a sharp cutoff of order unity. The dependence on $\alpha$ is most visible in the 
exponent of the stretched-exponential tail, which affects only the extreme, rare-event part of 
the distribution and has little impact on the bulk statistics. 
The same pattern will be encountered for the other statistical quantities studied in the following 
Sections.
This behavior contrasts with that of Poisson initial conditions in the Fr\'echet class studied in
\cite{Valageas2025}, corresponding to $\alpha < -3/2$ in Eq.(\ref{eq:lambda-psi}),
which lead to heavy power-law tails and therefore produce a much stronger dependence of the dynamics
on the exponent $\alpha$.

It is worth noting that, in exclusion-process-type transport models, integrated currents
across a bond can often be expressed in terms of Lagrangian displacements 
obeying hydrodynamical equations at the macroscopic level.
In our case, going back to dimensionfull variables and making explicit the time dependence
in the Lagrangian coordinate $q(x,t)$, we can define the current $j(x,t) = \rho v$,
the mass $M_>(t)$ to the right of the origin and the cumulative current $J(T)$ up to time $T$,
\be
j(x,t) = \rho v , \;\;\; M_>(t) = \int_0^\infty dx \, \rho(x,t) , \;\; 
J(T) = \int_0^T dt \, j(0,t) = M_>(T)-M_>(0) = - \rho_0 \, q(0,T) ,
\ee
where we used $q(0,0)=0$.
The mass $M_>(t)$ diverges because of the nonzero mean density as $x \to \infty$,
but the mass difference $M_>(T)-M_>(0)$ is finite and does not depend on the truncation of
the right boundary at distance $R$, with $R \to \infty$.
Thus, from the scaling laws (\ref{eq:re-scaling}), where we take $a=1$, we can see that the
probability distribution of the cumulative current is simply
\be
P_T(J) = \rho_0^{-1} T^{-(\alpha+1)/(2\alpha+3)} P_0\left( \rho_0^{-1} T^{-(\alpha+1)/(2\alpha+3)} J 
\right) ,
\ee
with the exponential tail
\be
|J| \gg T^{(\alpha+1)/(2\alpha+3)} : \;\; P_T(J) \sim  e^{-\Lambda_{\alpha} 2^{-\alpha-3/2}
\rho_0^{-(2\alpha+3)} T^{-(\alpha+1)} |J|^{2\alpha+3} } .
\ee
In contrast with the typical scalings $\propto T$ or $\propto \sqrt{T}$ of large-deviation
tails, the scaling $J \sim T^{(\alpha+1)/(2\alpha+3)}$ depends on the shape of the initial conditions
(\ref{eq:lambda-psi}). It directly follows from the self-similar dynamics (\ref{eq:re-scaling}),
associated with strongly non-Gaussian initial conditions.

\section{Two-point Eulerian distributions}
\label{sec:two-point}

We now turn to two-point statistics, which probe spatial correlations and therefore reveal 
how structures such as shocks and voids are organized in space. Unlike one-point quantities,
these observables capture the emergence of coherence and scaling behavior across different
length scales.

\subsection{Two first-contact parabolas}

In this Section we consider two-point Eulerian probability distributions, such as
$P_{x_1,x_2}(q_1,q_2)$, the probability that the Eulerian positions $x_1$ and $x_2$ map to
the Lagrangian coordinates $q_1$ and $q_2$. 
As in Section~\ref{sec:one-point-Eulerian}, we first examine the probability
$P_{x_1,x_2}(q_1,c_1;q_2,c_2)$ that the first-contact points lie at $q_1$ and $q_2$,
with parabola heights $c_1$ and $c_2$.
For the Poisson point process (\ref{eq:Lambda-Psi0}), this is
\bea
x_1 < x_2 : &&  P_{x_1,x_2}(q_1,c_1;q_2,c_2) = \biggl [ \frac{\lambda(\psi_\star)}{x_2-x_1} 
\delta_D(q_1-q_\star) \delta_D(q_2-q_\star) + \theta(q_1<q_\star) \theta(q_2 > q_\star)
\lambda( c_1 + (q_1-x_1)^2/2 ) 
\nonumber \\
&& \times \lambda( c_2 + (q_2-x_2)^2/2) \biggl ] 
e^{-\int_{-\infty}^{q_\star} dq 
\int_0^{c_1-(q-x_1)^2/2} d\psi \, \lambda(\psi) - \int_{q_\star}^{\infty} dq 
\int_0^{c_2-(q-x_2)^2/2} d\psi \, \lambda(\psi) } ,
\label{eq:Pq1c1q2c2}
\eea
where $\delta_D$ and $\theta$ denote the Dirac delta distribution and Heaviside function, 
and $(q_\star,\psi_\star)$ is the intersection of the two parabolas:
\be
q_\star = \frac{x_1+x_2}{2} + \frac{c_1-c_2}{x_2-x_1} , \;\;\;
\psi_\star = c_1 - \frac{(q_\star-x_1)^2}{2} = c_2 - \frac{(q_\star-x_2)^2}{2} 
= \frac{c_1+c_2}{2} - \frac{(x_2-x_1)^2}{8} - \frac{(c_2-c_1)^2}{2 (x_2-x_1)^2} .
\ee
For $q<q_\star$, the parabola ${\cal P}_{x_1,c_1}$ lies above ${\cal P}_{x_2,c_2}$,
while for $q>q_\star$ the opposite holds, since we take $x_1 < x_2$.
The first term in the bracket corresponds to the case where both parabolas share a common 
first-contact point with the initial potential $\psi_0(q)$, which is also their intersection
$(q_\star,\psi_\star)$.
The second term corresponds to the case where the two first-contact points $q_1$ and $q_2$ are
distinct, with $q_1<q_\star<q_2$.
Thus, the key idea is to generalize the one-parabola first-contact construction to pairs of 
parabolas, which allows us to encode the joint behavior of the velocity potential at two positions.
The resulting expressions reduce the two-point problem to integrals over a handful of variables describing the relative configuration of the relevant Poisson points and parabolas.
We now introduce the change of variables
\be
\bar x  = \frac{x_1+x_2}{2} , \;\;\; x = x_2 - x_1 > 0 , \;\;\;  q_1 = \bar x + q'_1 , \;\;\; 
q_2 = \bar x + q'_2 ,
\label{eq:qp1-qp2-def}
\ee
integrate over $(c_1,c_2)$, and transform the integration variables from $(c_1,c_2)$ to 
$(q_\star',\psi_\star)$. This yields
\be
P_x(q'_1,q'_2) = \! \int_{-\infty}^{\infty} \!\! dq'_\star \int_{\psi_{\min}(q'_\star)}^{\infty} \!\!\! 
d\psi_\star \, \bigg[ \lambda(\psi_\star) \delta_D(q'_1-q'_\star) \delta_D(q'_2-q'_\star) 
+ x \theta(q'_1<q'_\star) \theta(q'_2 > q'_\star) \psi_-(q'_1)^{\alpha} 
\psi_+(q'_2)^{\alpha} \bigg] e^{-{\cal I}} ,
\label{eq:Pq1q2}
\ee
with
\bea
&& \psi_-(q') = \max\left[ 0 , \psi_\star + \frac{1}{2} \left( q'_\star + \frac{x}{2} \right)^2 - \frac{1}{2}
\left ( q' + \frac{x}{2} \right)^2 \right] , \;\;\;\;
\psi_+(q') = \max \left[ 0 , \psi_\star + \frac{1}{2} \left( q'_\star - \frac{x}{2} \right)^2 - \frac{1}{2}
\left ( q' - \frac{x}{2} \right)^2 \right] , \nonumber \\
&& {\cal I}(\psi_\star,q'_\star) = \frac{1}{\alpha+1} \left[ 
\int_{-\infty}^{q'_\star} dq' \psi_-(q')^{\alpha+1} 
+ \int_{q'_\star}^{\infty} dq' \psi_+(q')^{\alpha+1} \right] , \;\;\;
\eea
and
\be
| q'_\star | \geq \frac{x}{2} : \;\; \psi_{\min}(q'_\star) = 0 , \;\;\; | q'_\star | < \frac{x}{2} : \;\; 
\psi_{\min}(q'_\star) = - \frac{1}{2} \left( | q'_\star | - \frac{x}{2} \right)^2 .
\label{eq:psi_min}
\ee
The minimum $\psi_{\min}$ enforces the conditions that the parabola heights $c_1$ and $c_2$
remain positive and that the upper portions of the parabolas ${\cal P}_1$ and ${\cal P}_2$ are not
entirely below one another, ensuring that each has a non-vanishing interval of admissible
first-contact points.
For the first term in Eq.(\ref{eq:Pq1q2}), the factor $\lambda(\psi_\star)$ actually implies
$\psi_\star > 0$, since this configuration requires a Poisson point 
at the intersection $(q_\star,\psi_\star)$.
This expression shows explicitly that $P_x(q'_1,q'_2)$ depends only on the separation $x=x_2-x_1$, 
owing to translational invariance and the centered coordinates in (\ref{eq:qp1-qp2-def}).
By parity symmetry, we also have ${\cal I}(\psi_\star,-q'_\star)={\cal I}(\psi_\star,q'_\star)$.
For $x=0$, the two parabolas coincide, and the quantity ${\cal I}$ reduces to the expression found
in Eq.(\ref{eq:P_0-q}) for the one-point distribution:
\be
x=0 : \;\;\; {\cal I}(\psi_\star,q'_\star) = \Lambda_{\alpha} 
\left( \psi_\star + q'_\star \, \! ^{2}/2 \right)^{\alpha+3/2} .
\ee

For later use we define the quantities ${\cal A}_{\nu}(x,q'_\star)$
and ${\cal R}_{\nu}(x) $,
\be
\nu > -1 , \;\; x \geq 0 : \;\; {\cal A}_{\nu}(x,q'_\star) = \int_0^{\infty} d \psi_\star \, 
\psi_\star^{\nu} e^{- {\cal I}(\psi_\star,q'_\star) } , 
\;\;\; {\cal R}_{\nu}(x) = \int_{-\infty}^{\infty} d q'_\star \, {\cal A}_{\nu}(x,q'_\star) ,
\label{eq:A-nu-def}
\ee
which for $x=0$ give
\be
{\cal R}_{\nu}(0) = \frac{2\sqrt{2\pi} \Gamma(\nu+1)}{(2\alpha+3) \Gamma(\nu+3/2)} \,
\Gamma\left(\frac{2\nu+3}{2\alpha+3}\right) \Lambda_{\alpha}^{-(2\nu+3)/(2\alpha+3)} ,
\label{eq:R-nu-0}
\ee
\be
{\cal R}_{\nu}'(0) = - \frac{\sqrt{2\pi} \Gamma(\alpha+\nu+3) 
\Gamma\left(\frac{2\alpha+2\nu+5}{2\alpha+3}\right) } {(\alpha+1) (\alpha+3/2) (\nu+1) 
\Gamma(\alpha+\nu+5/2)} \Lambda_{\alpha}^{-(2\alpha+2\nu+5)/(2\alpha+3)} ,
\label{eq:Rp-nu-0}
\ee
and for large separations,
\be
x \gg 1 : \;\;\; {\cal R}_{\nu}(x) \sim x^{- (\nu+3/2) (2\alpha+1) } \, e^{- \Lambda_{\alpha} 2^{-3\alpha-7/2}
x^{2\alpha+3} } .
\label{eq:R-nu-large-x}
\ee

\subsection{Void probabilities}
\label{sec:Void-probabilities}

\subsubsection{Probability of an empty interval}

\begin{figure}
\centering
\includegraphics[height=6cm,width=0.48\textwidth]{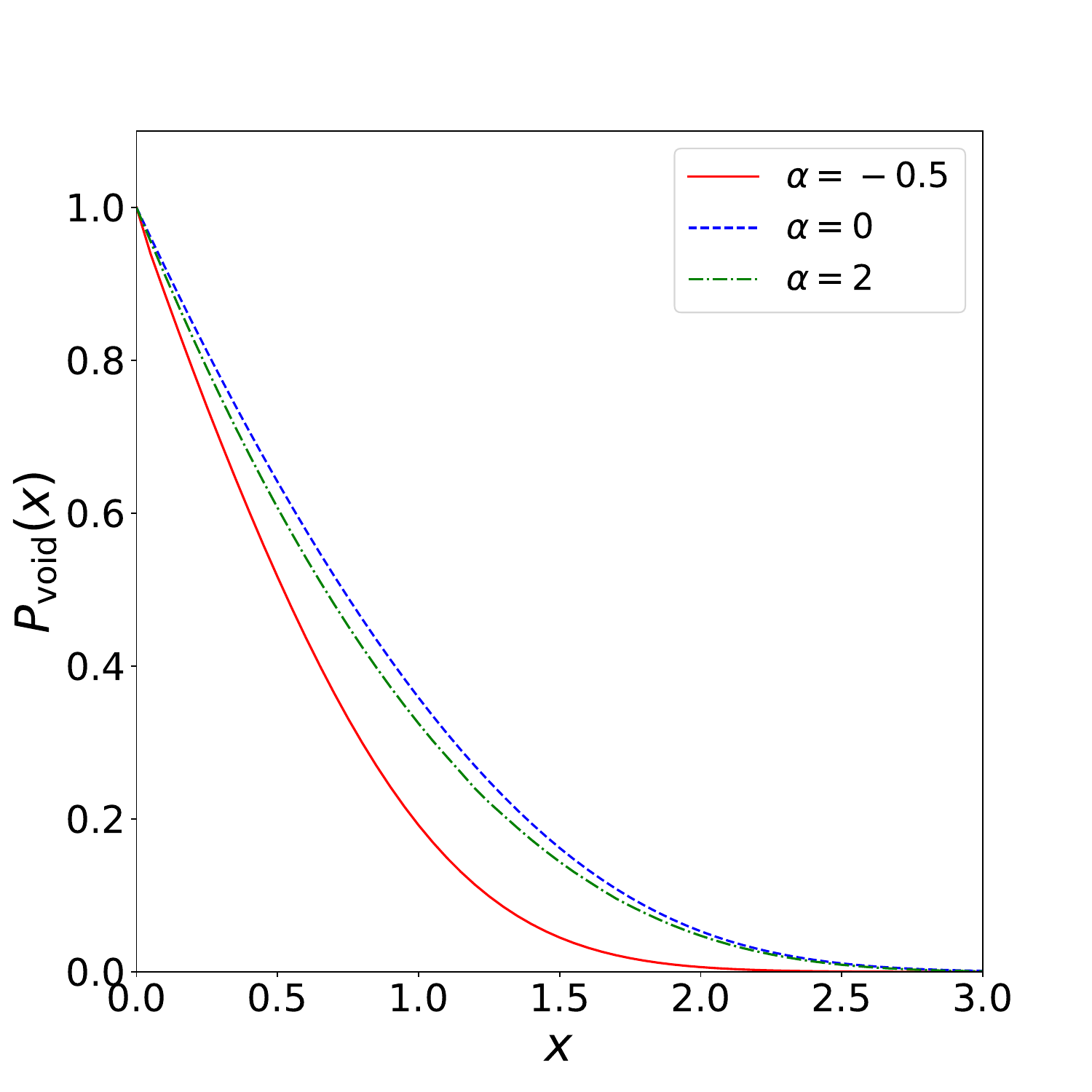}
\includegraphics[height=6cm,width=0.48\textwidth]{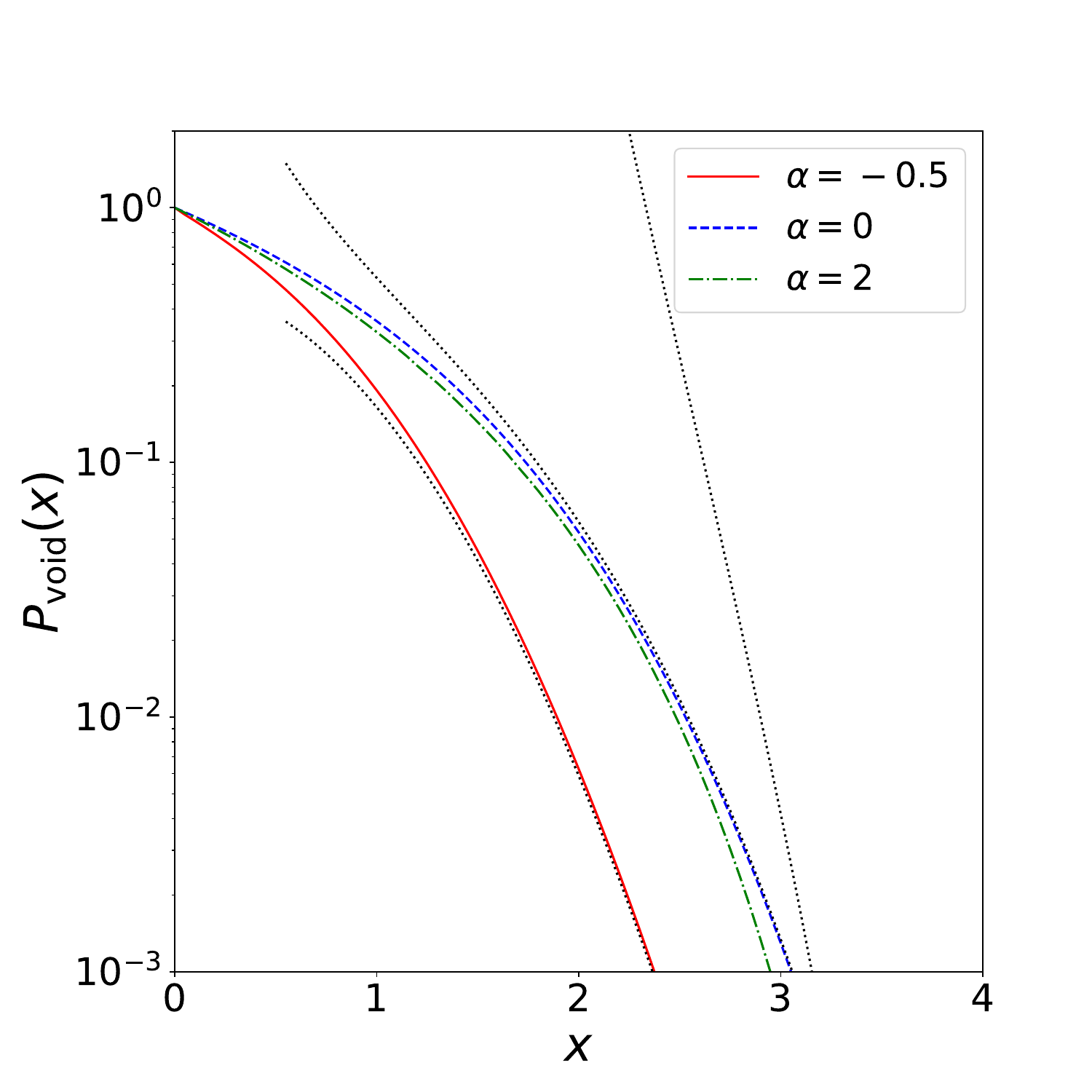}
\caption{
Void probability $P_{\rm void}(x)$ from Eq.(\ref{eq:Pvoid}), for the cases 
$\alpha=-1/2, 0$, and $2$, as in Fig.~\ref{fig:P0_q}.
The dotted lines in the right panel are the asymptotic stretched exponentials 
(\ref{eq:Pvoid-x-0-large-x}).
}
\label{fig:Pvoid}
\end{figure}

A particularly transparent observable in this framework is the probability that a given
Eulerian interval is completely empty, i.e. contains no mass.
This "void probability" directly measures how the system partitions space into empty regions and 
shocks, and it will also serve as a building block for the void multiplicity function
studied in Section~\ref{sec:multiplicity-voids} below.
The overdensity within the Eulerian interval $[x_1,x_2]$ is defined by 
\be
\rho_{x_1,x_2} = \frac{q_2-q_1}{x_2-x_1} \geq 0 ,
\ee
where the density has been rescaled by the mean density $\rho_0$, though we retain
the simpler notation $\rho$ in the following.
If the two parabolas share the same contact point, $q_1=q_2=q_\star$, then the density vanishes,
$\rho=0$, and the interval $[x_1,x_2]$ contains no matter. From Eq.(\ref{eq:Pq1q2}), the probability 
that the interval is empty is therefore
\be
P_{\rm void}(x) = {\cal R}_{\alpha}(x) ,
\label{eq:Pvoid}
\ee
where the function ${\cal R}_{\alpha}(x)$ was introduced in Eq.(\ref{eq:A-nu-def}).
Using the results (\ref{eq:R-nu-0}) and (\ref{eq:R-nu-large-x}), we obtain
\be
P_{\rm void}(0) = 1  \;\;\; \mbox{and for} \;\;\; x \gg 1 : \;\;\; 
P_{\rm void}(x) \sim x^{-(\alpha+3/2) (2\alpha+1)} \, 
e^{-\Lambda_\alpha 2^{-3\alpha-7/2} x^{2\alpha+3} } .
\label{eq:Pvoid-x-0-large-x}
\ee
Thus, the void probability approaches unity as $x\to 0$ and decreases as a stretched exponential 
for large intervals. 
Its exponent encodes how sensitive the system is to the tail of the initial potential,
and provides a compact quantitative measure of intermittency in the spatial distribution
of matter.
The result $P_{\rm void}(0) = 1$ reflects the fact that voids fill the entire
Eulerian space, with matter concentrated in Dirac-like density peaks of zero width.
The void probability $P_{\rm void}(x)$ is shown in Fig.~\ref{fig:Pvoid}.
it displays a monotonic decrease from unity for $x>0$ with stretched-exponential tails that agree
with Eq.(\ref{eq:Pvoid-x-0-large-x}).

\subsubsection{Multiplicity function of voids and distance between shocks}
\label{sec:multiplicity-voids}

\begin{figure}
\centering
\includegraphics[height=6.cm,width=0.33\textwidth]{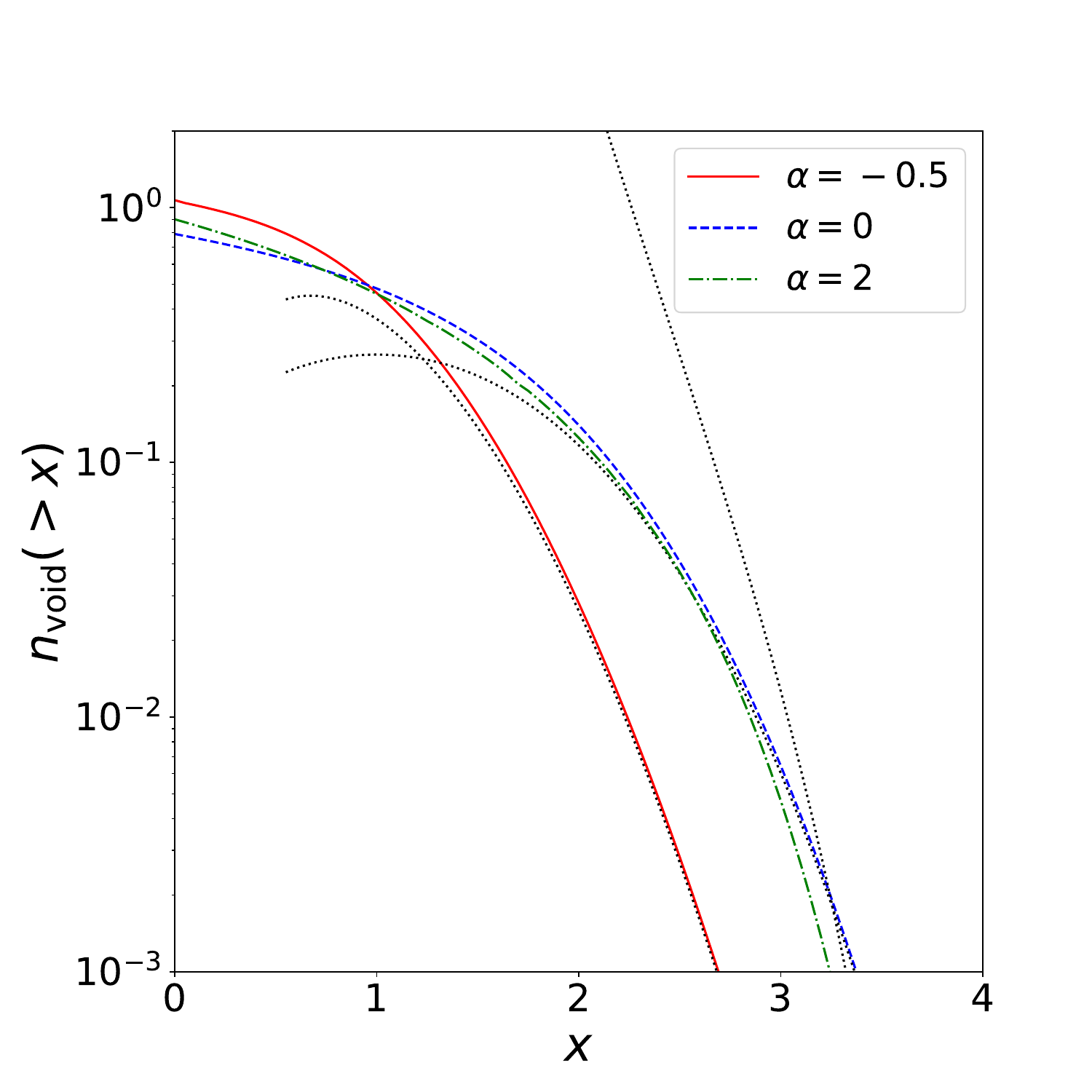}
\includegraphics[height=6.cm,width=0.33\textwidth]{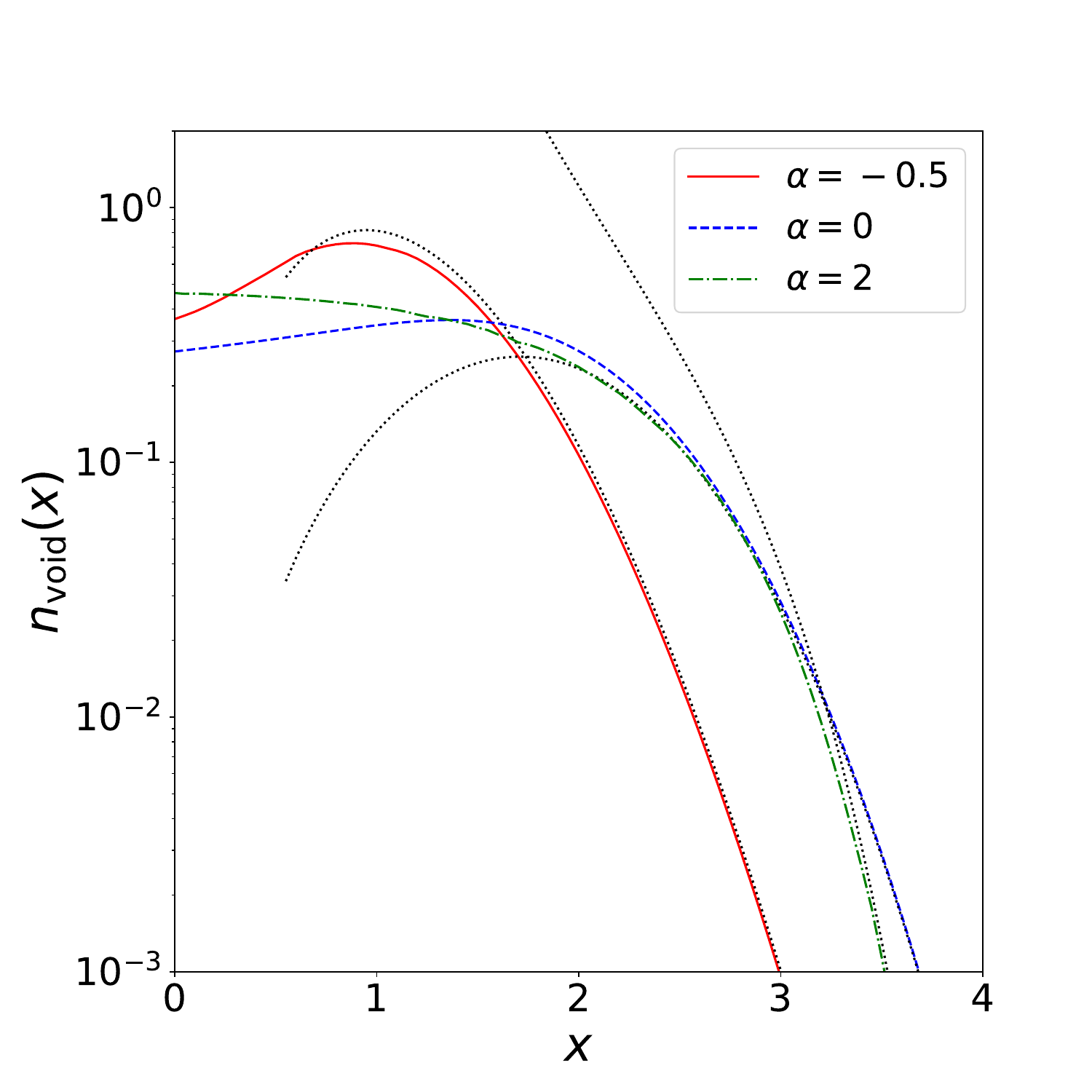}
\includegraphics[height=6.cm,width=0.32\textwidth]{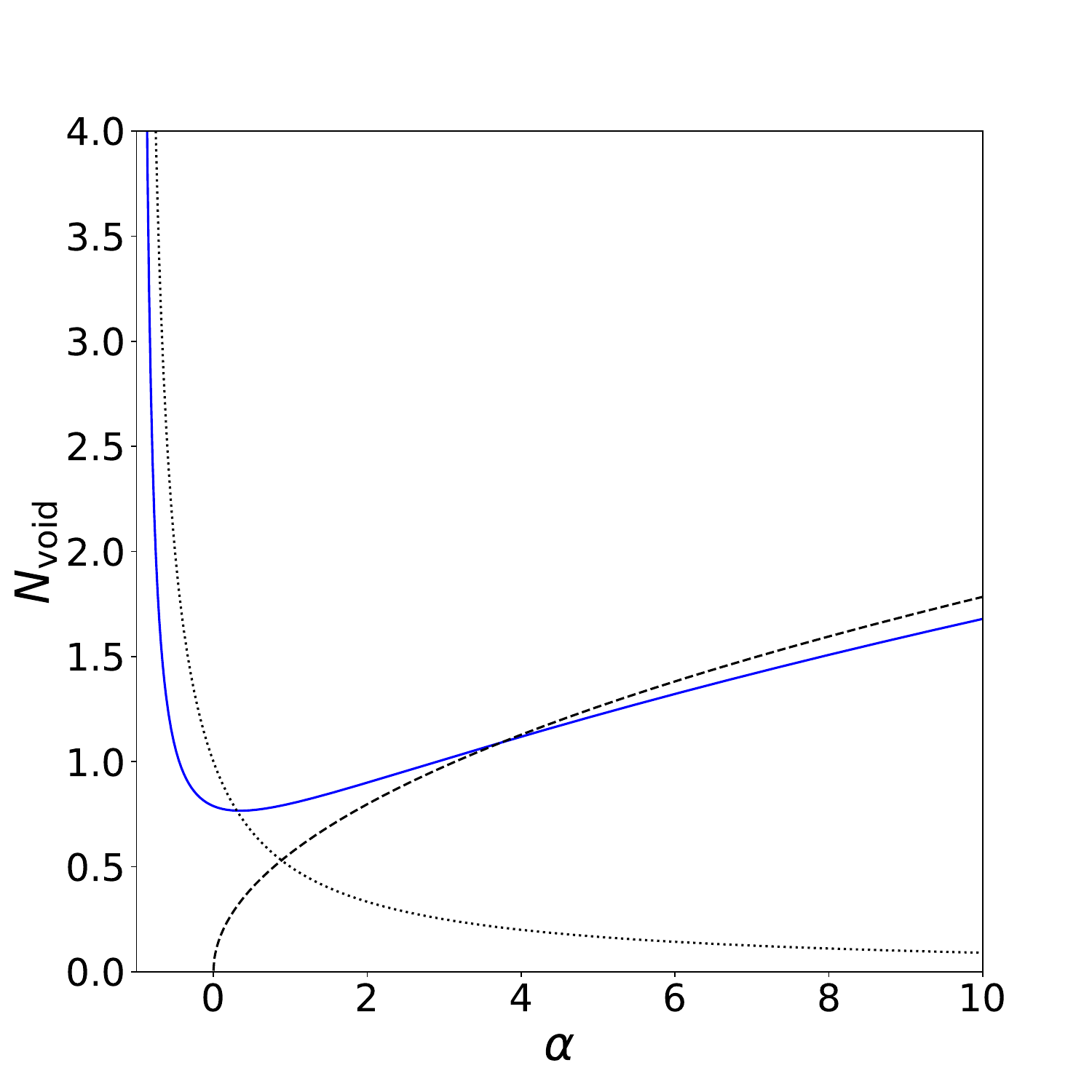}
\caption{
{\it Left panel:} cumulative void multiplicity function $n_{\rm void}(>x)$ from 
Eq.(\ref{eq:n-void-R-alpha}). 
{\it Middle panel:} void multiplicity function $n_{\rm void}(x)$ from 
Eq.(\ref{eq:n-void-R-alpha}). 
{\it Right panel:} number density of voids $N_{\rm void}$ as a function of $\alpha$ (solid blue line).
In the left and middle panels, the dotted lines are the stretched exponentials associated with
Eq.(\ref{eq:nvoid-power-law}).
In the right panel the dotted and dashed lines are the asymptotic regimes (\ref{eq:Nvoid-asymp}).
}
\label{fig:nvoid}
\end{figure}

The void multiplicity function $n_{\rm void}(x)$ expresses the information contained in
$P_{\rm void}(x)$ by specifying how many voids of a given size exist per unit length.
It is analogous to a size distribution of domains in coarsening systems, or to a halo mass function
in cosmology, and it describes how shocks carve up space.
Thus, let $n_{\rm void}(x) dx$ denote the number of voids per unit length with sizes in the interval 
$[x,x+dx]$. It is related to the void probability $P_{\rm void}(x)$ by
\be
P_{\rm void}(x) = \int_x^{\infty} dx' n_{\rm void}(x') \, (x'-x) , \;\;\; \mbox{whence} \;\;\;
n_{\rm void}(>x) = - \frac{dP_{\rm void}}{dx} = - {\cal R}'_{\alpha}(x) , \;\;
n_{\rm void}(x) = \frac{d^2P_{\rm void}}{dx^2} = {\cal R}''_{\alpha}(x) .
\label{eq:n-void-R-alpha}
\ee
Using the asymptotic expression (\ref{eq:R-nu-large-x}) and $P_{\rm void}(0)=1$, we obtain
\be
x \gg 1 : \;\; n_{\rm void}(x) \sim x^{-2\alpha^2+5/2} \, e^{-\Lambda_\alpha 2^{-3\alpha-7/2} x^{2\alpha+3} } , 
\;\;\; \mbox{and} \;\; \int_0^{\infty} dx \, n_{\rm void}(x) \, x = 1 ,
\label{eq:nvoid-power-law}
\ee
which again reflects the fact that voids occupy the entire Eulerian space,
whereas all mass is compressed into shocks of zero width.
From a physical standpoint, it emphasizes that the relevant structures are the shocks themselves
and their separation, while the voids form the dominant background.
From Eq.(\ref{eq:Rp-nu-0}) the number of voids $N_{\rm void}$ per unit length is
\be
N_{\rm void} = n_{\rm void}(>0) = - {\cal R}'_{\alpha}(0) = \frac{ \sqrt{2\pi} \Gamma(2\alpha+3) 
\Gamma\left( \frac{4\alpha+5}{2\alpha+3} \right) } { (\alpha+1)^2 (\alpha+3/2) \Gamma(2\alpha+5/2) }
\Lambda_{\alpha}^{-(4\alpha+5)/(2\alpha+3)} .
\label{eq:Nvoid}
\ee
Its limiting behaviors are
\be
\alpha \to -1 : \;\; N_{\rm void} \sim \frac{1}{\alpha+1} , \;\;\;
\mbox{and for} \;\;\; \alpha \to \infty : \;\; 
N_{\rm void} \sim \sqrt{\alpha/\pi} ,
\label{eq:Nvoid-asymp}
\ee
so that the number of voids per unit length diverges in both limits $\alpha \to -1$ and 
$\alpha \to \infty$.
The mean void size is given by
\be
\langle x \rangle_{\rm void} = \frac{ \int_0^{\infty} dx \, n_{\rm void}(x) \, x }{ \int_0^{\infty} dx \, n_{\rm void}(x)}
= \frac{1}{N_{\rm void}} , 
\ee
and therefore vanishes as $\alpha \to -1$ and $\alpha \to \infty$.
Since the system consists of a series of shocks separated by voids, the void multiplicity function
$n_{\rm void}(x)$ also provides the probability distribution $P(x_s)$ of the distance $x_s$ between
adjacent shocks,
\be
P(x_s) = \frac{n_{\rm void}(x_s)}{N_{\rm void}} = - \frac{{\cal R}''_{\alpha}(x_s)}{{\cal R}'_{\alpha}(0)} .
\label{eq:P-xs}
\ee
The void multiplicity functions $n_{\rm void}(>x)$ and $n_{\rm void}(x)$ are shown 
in the left and central panels of Fig.~\ref{fig:nvoid}, and the mean number of voids per unit length
is displayed in the right panel.
The void multiplicity function $n_{\rm void}(x)$ is somewhat more sensitive to $\alpha$ than
$P_{\rm void}(x)$ as the sign of its slope at the origin can change sign with the value of
$\alpha$.

\subsection{Two-point velocity correlation and energy power spectrum}
\label{sec:energy-spectrum}

\begin{figure}
\centering
\includegraphics[height=6cm,width=0.48\textwidth]{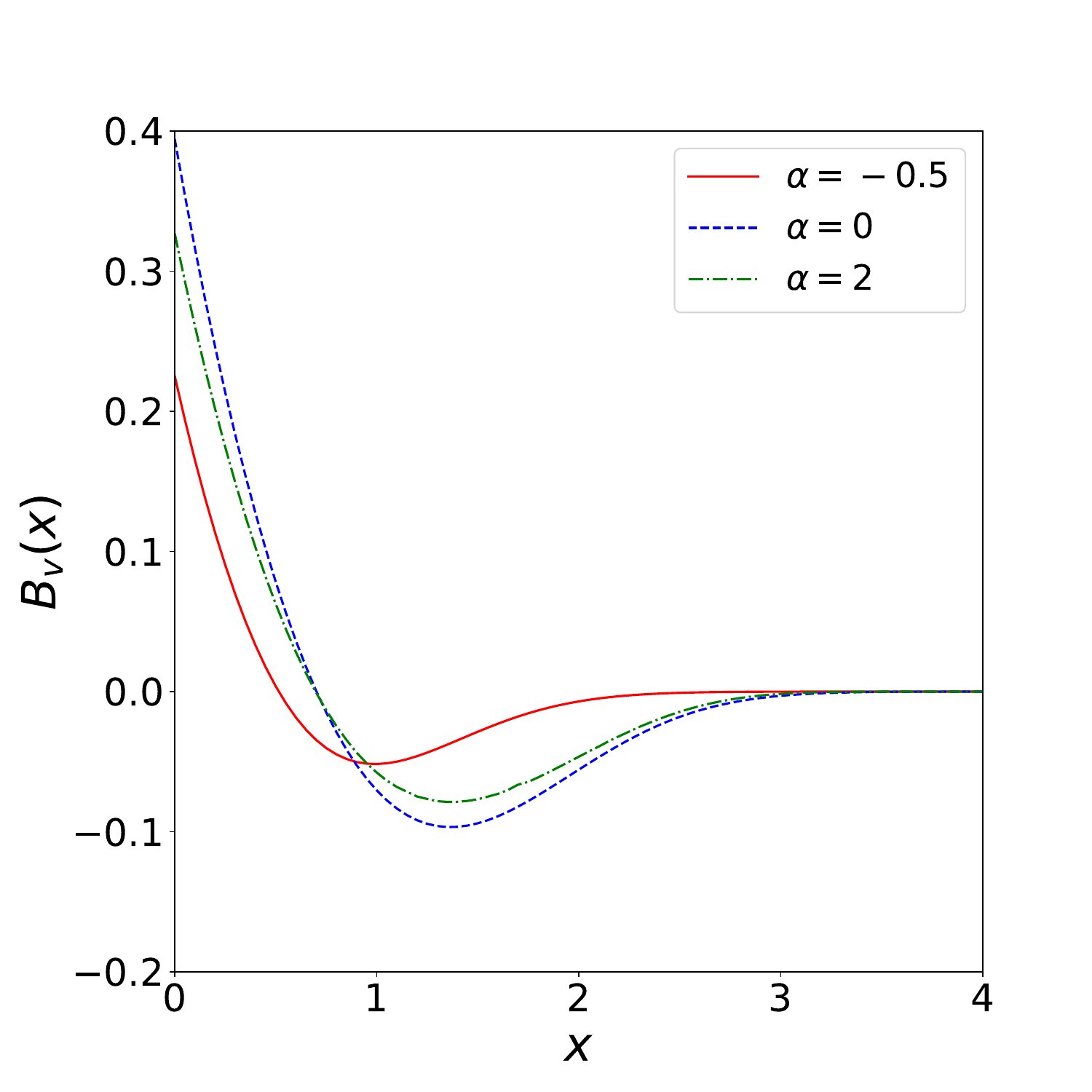}
\includegraphics[height=6cm,width=0.48\textwidth]{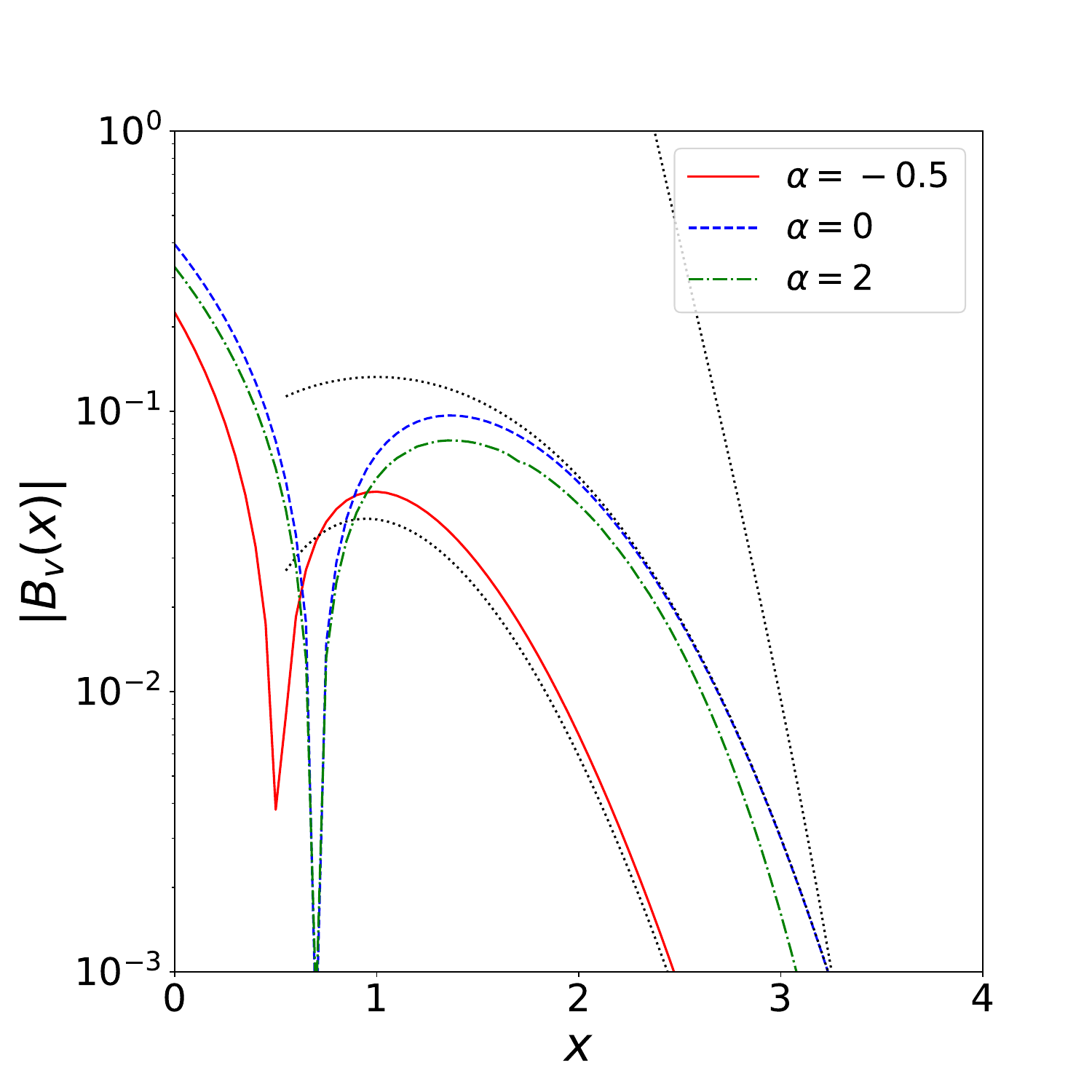}
\caption{
Velocity correlation $B_v(x)$ for  the cases $\alpha= -0.5, 0$, and $2$.
In the right panel the dotted lines are the asymptotic stretched exponentials (\ref{eq:Bv-asymp}).
}
\label{fig:Bv}
\end{figure}

\begin{figure}
\centering
\includegraphics[height=6cm,width=0.48\textwidth]{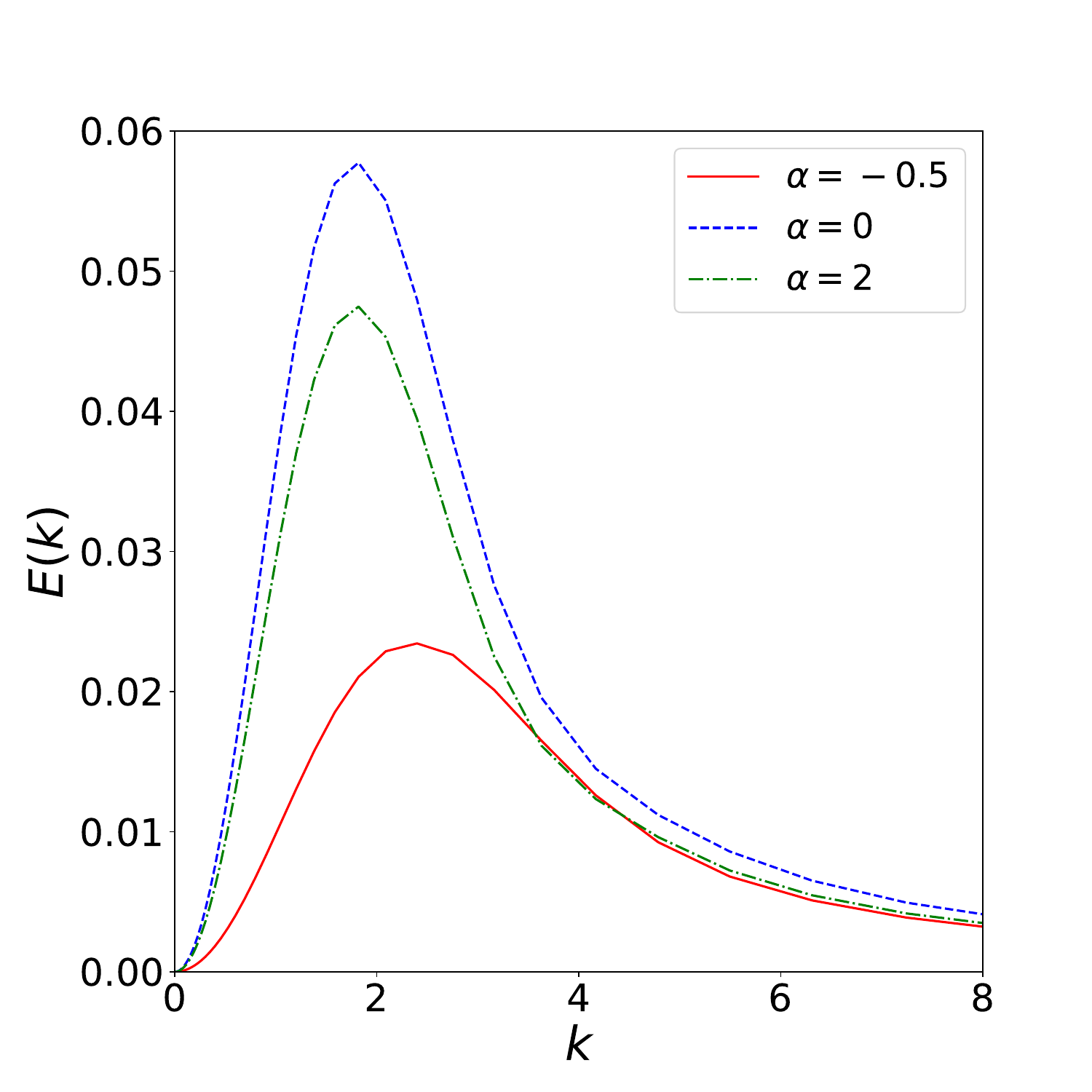}
\includegraphics[height=6cm,width=0.48\textwidth]{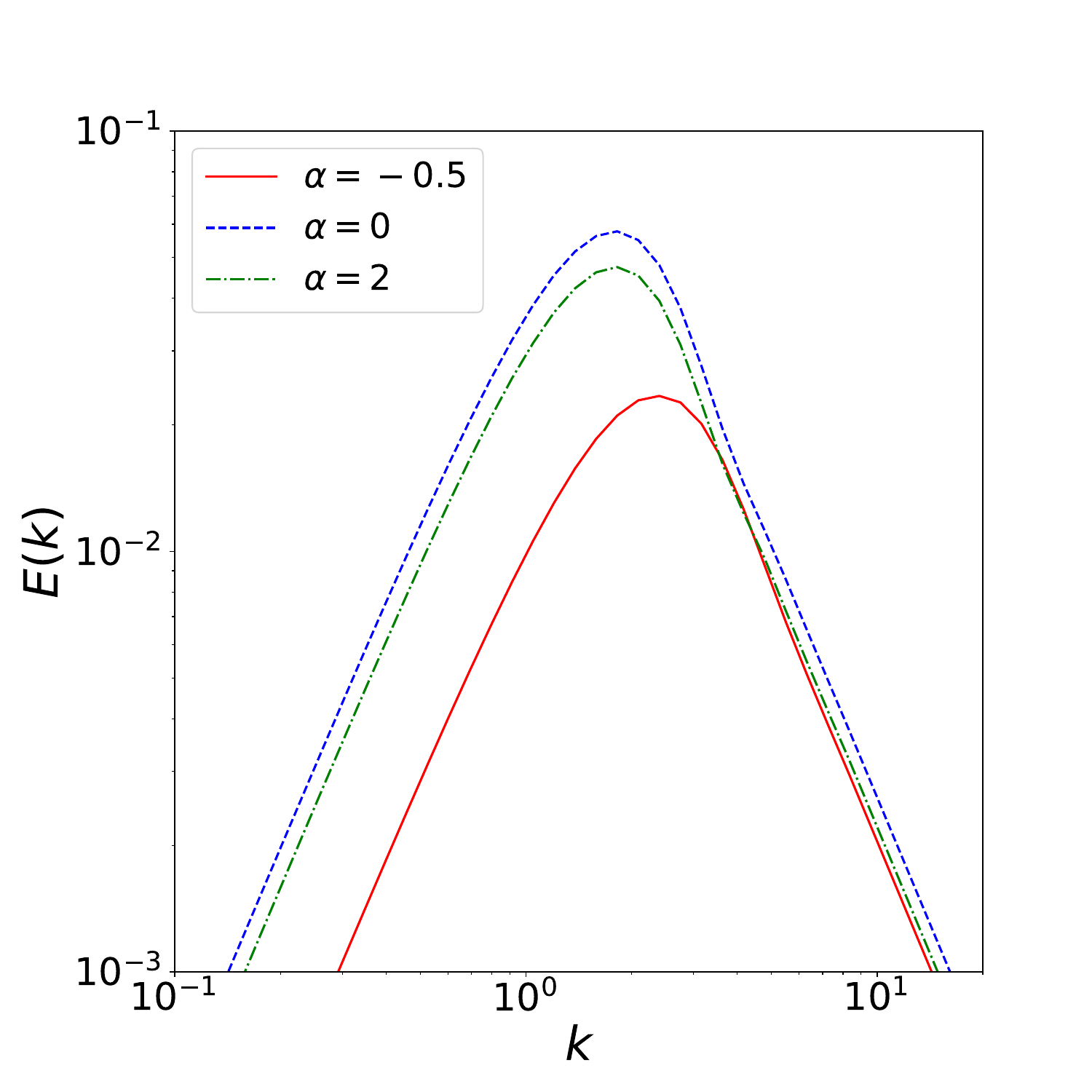}
\caption{
Velocity power spectrum $E(k)$ on linear and logarithmic scales.
}
\label{fig:Ek}
\end{figure}

We now turn to the velocity and density correlation functions. Together with their Fourier
counterparts (power spectra), they summarize how fluctuations at different scales are coupled
by the dynamics.
Their asymptotic behavior at small and large separations reveals how shocks, voids, and the underlying Poisson statistics shape the flow across scales.
Moreover, in many applications, such as turbulence or cosmology, these correlation functions and
power spectra are the primary observables used to characterize the system.
Using $v_1=x_1-q_1=-q'_1-x/2$, $v_2=x_2-q_2=-q'_2+x/2$, and the expression (\ref{eq:Pq1q2})
for the joint distribution of $q'_1$ and $q'_2$, we obtain the velocity correlation
between two points separated by $x=x_2-x_1 \geq 0$,
\be
B_v(x) \equiv \langle v_1 v_2 \rangle_x = \int_{-\infty}^{\infty} dq'_\star 
\int_0^{\infty} d \psi_\star \left[ \psi_\star^{\alpha} 
\left( q'_\star \, \! ^2 - \frac{x^2}{4} \right) - \frac{x}{(\alpha+1)^2} \psi_\star^{2\alpha+2} 
\right]  e^{- {\cal I}(\psi_\star,q'_\star) } ,
\label{eq:v1v2-1}
\ee
where we have already integrated over $q'_1$ and $q'_2$.
Configurations with $\psi_\star \leq 0 $ do not contribute, as in this case
the parabolic arcs ${\cal P}_1$ and ${\cal P}_2$ in the upper half-plane are disjoints.
Then, the integral over either symmetric arc yields a vanishing mean velocity:
the contributions from the two sides of $x_1$ along ${\cal P}_1$ have identical weights
but opposite velocities $v_1 = x_1-q_1$. 
As for the Fr\'echet-type case studied in \cite{Valageas2025}, an integration by parts with 
respect to $\psi_\star$ shows that (\ref{eq:v1v2-1}) can also be written as
\be
B_v(x) = \frac{1}{\alpha+1} \int_{-\infty}^{\infty} dq'_\star \left( x \frac{\partial}{\partial x} 
- q'_\star \frac{\partial}{\partial q'_\star} \right) {\cal A}_{\alpha+1} ,
\ee
where ${\cal A}_{\nu}$ is defined in Eq.(\ref{eq:A-nu-def}).
Performing the integration by parts over $q'_\star$ yields
\be
x \geq 0 : \;\;\; B_v(x) =  \frac{1}{\alpha+1}  \frac{d}{dx} \left[ x {\cal R}_{\alpha+1}(x) \right] , 
\label{eq:Bv-deriv}
\ee
with ${\cal R}_{\nu}$ defined in Eq.(\ref{eq:A-nu-def}).
By parity, $B_v(-x) = B_v(x) = B_v(|x|)$.
This gives the small- and large-scale asymptotic behaviors
\be
|x| \ll 1 : B_v(x) = \frac{{\cal R}_{\alpha+1}(0)}{\alpha+1} 
+ \frac{2 {\cal R}'_{\alpha+1}(0)}{\alpha+1} |x| + \dots ,   \;\;\; |x| \gg 1 :  \;\;\;  
B_v(x) \sim -  x^{-2\alpha^2-4\alpha+1/2} \, e^{-\Lambda_\alpha 2^{-3\alpha-7/2} x^{2\alpha+3} } .
\label{eq:Bv-asymp}
\ee
The non-analytic $|x|$ term, which carries a negative prefactor, reflects the contribution of shocks. 
Its linear dependence on $|x|$ corresponds to the probability of encountering at least one shock
within an interval of size $|x|$, which decreases linearly following the slope of the complementary
void probability $1-P_{\rm void}(|x|)$.
A shock at position $x_s$ induces a discontinuous decrease in the velocity, 
$v(x_s^+)-v(x_s^-)<0$, producing the characteristic sawtooth pattern of Burgers dynamics,
illustrated in Fig.~\ref{fig:realization}.
This explains the negative coefficient of the $|x|$ term in (\ref{eq:Bv-asymp}).
The stretched-exponential decay of correlations at large separations is consistent with the 
behavior of the void probability and one-point distributions, and it confirms that the Weibull class 
produces short-ranged correlations despite strong local intermittency.
This provides a coherent picture in which shocks dominate small-scale structure, while large scales 
remain only weakly correlated.

The resulting velocity correlation function is shown in Fig.~\ref{fig:Bv}.
It is positive at small separations, $|x| \lesssim 1$, and it becomes negative at large separations, 
$|x|  \gg 1$, in agreement with the asymptotics (\ref{eq:Bv-asymp}) and Eq.(\ref{eq:R-nu-0}).

The energy spectrum $E(k)$ is the Fourier transform of $B_v$,
\be
E(k) \equiv \int_{-\infty}^{\infty} \frac{dx}{2\pi} B_v(x) e^{ikx} 
= \int_0^{\infty} \frac{dx}{\pi} B_v(x) \cos(kx) =  
\frac{k}{\pi (\alpha+1)} \int_0^{\infty} dx \, x {\cal R}_{\alpha+1}(x) \sin(kx) ,
\label{eq:Ek-def}
\ee
where we used $B_v(-x) = B_v(x)$ and Eq.(\ref{eq:Bv-deriv}), followed by an integration by parts.
This implies the low-wavenumber behavior
\be
| k | \ll 1 : \;\; E(k) \simeq \frac{k^2}{\pi (\alpha+1)} 
\int_0^{\infty} dx \, x^2 {\cal R}_{\alpha+1}(x) \propto k^2 ,
\label{eq:Ek-small-k-power}
\ee
so that the energy spectrum universally exhibits a quadratic decay at small $k$.
At large wavenumbers, the non-analytic $|x|$ term in $B_v(x)$ produces the universal $k^{-2}$ 
tail
\be
| k | \gg 1 : \;\;\; E(k) \simeq \sqrt{\frac{2}{\pi}}  
\frac{8 \Gamma(2\alpha+2) \Gamma\left(\frac{4\alpha+7}{2\alpha+3}\right)} 
{(\alpha+1) (\alpha+2) \Gamma(2\alpha+7/2)} \Lambda_{\alpha}^{-(4 \alpha+7)/(2\alpha+3)} k^{-2} .
\label{eq:Ek-large-k-power}
\ee
The resulting energy spectrum is shown in Fig.~\ref{fig:Ek}. It remains positive for all wavenumbers
and matches the asymptotic behaviors (\ref{eq:Ek-small-k-power})-(\ref{eq:Ek-large-k-power}).

From a broader perspective, the results obtained above illustrate how spatial correlations are 
generated dynamically from initially uncorrelated configurations through nonlinear transport. This 
mechanism is directly relevant to exclusion processes and related interacting particle systems, where 
correlations emerge from the interplay between initial fluctuations and deterministic hydrodynamic 
evolution \cite{Doyon2023}.
This suggests that stretched-exponential tails could also appear in these systems,
for strongly non-Gaussian initial conditions where extreme fluctuations are associated with
atypical configurations.

\subsection{Density correlation and power spectrum}
\label{sec:Density-field}

\begin{figure}
\centering
\includegraphics[height=6.cm,width=0.33\textwidth]{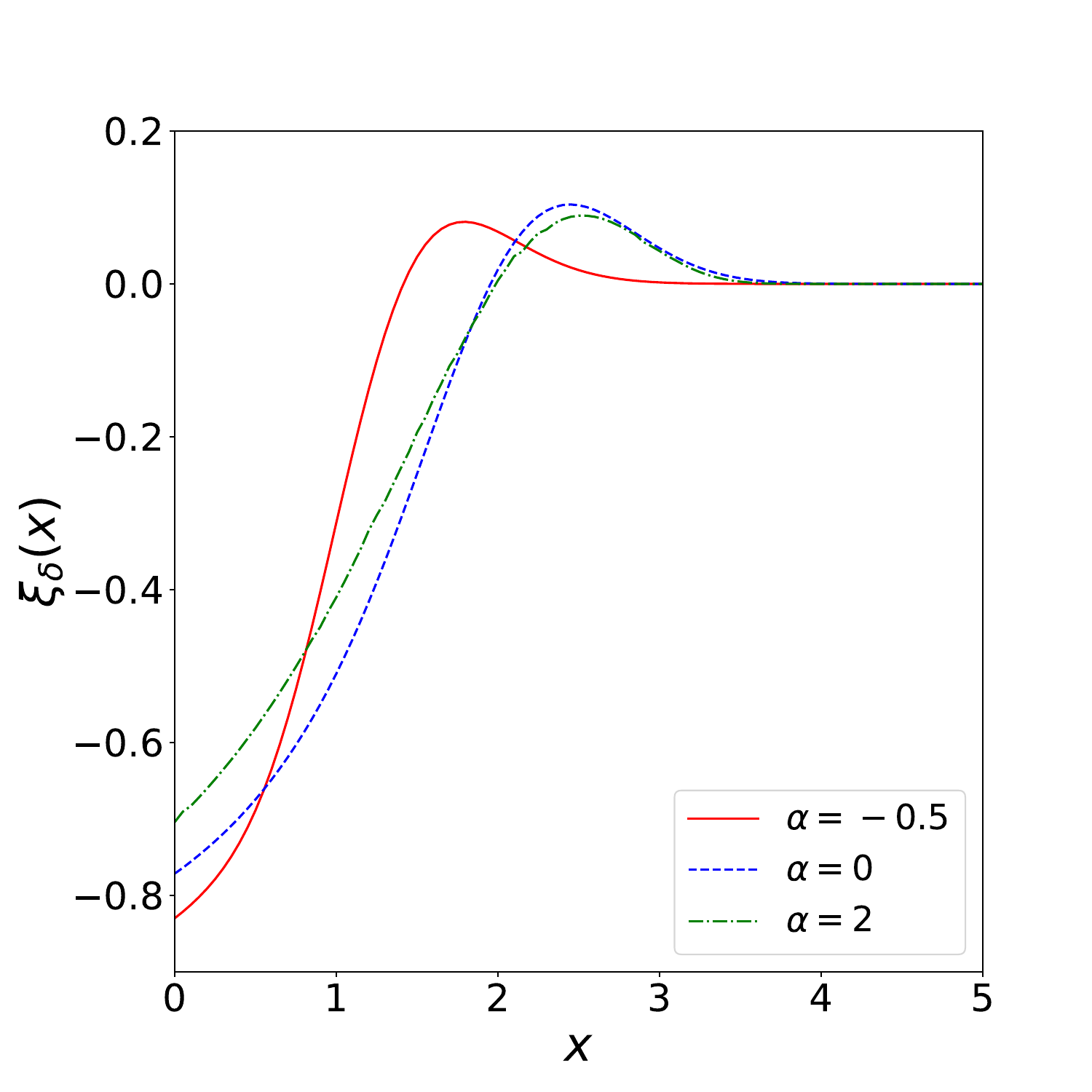}
\includegraphics[height=6.cm,width=0.33\textwidth]{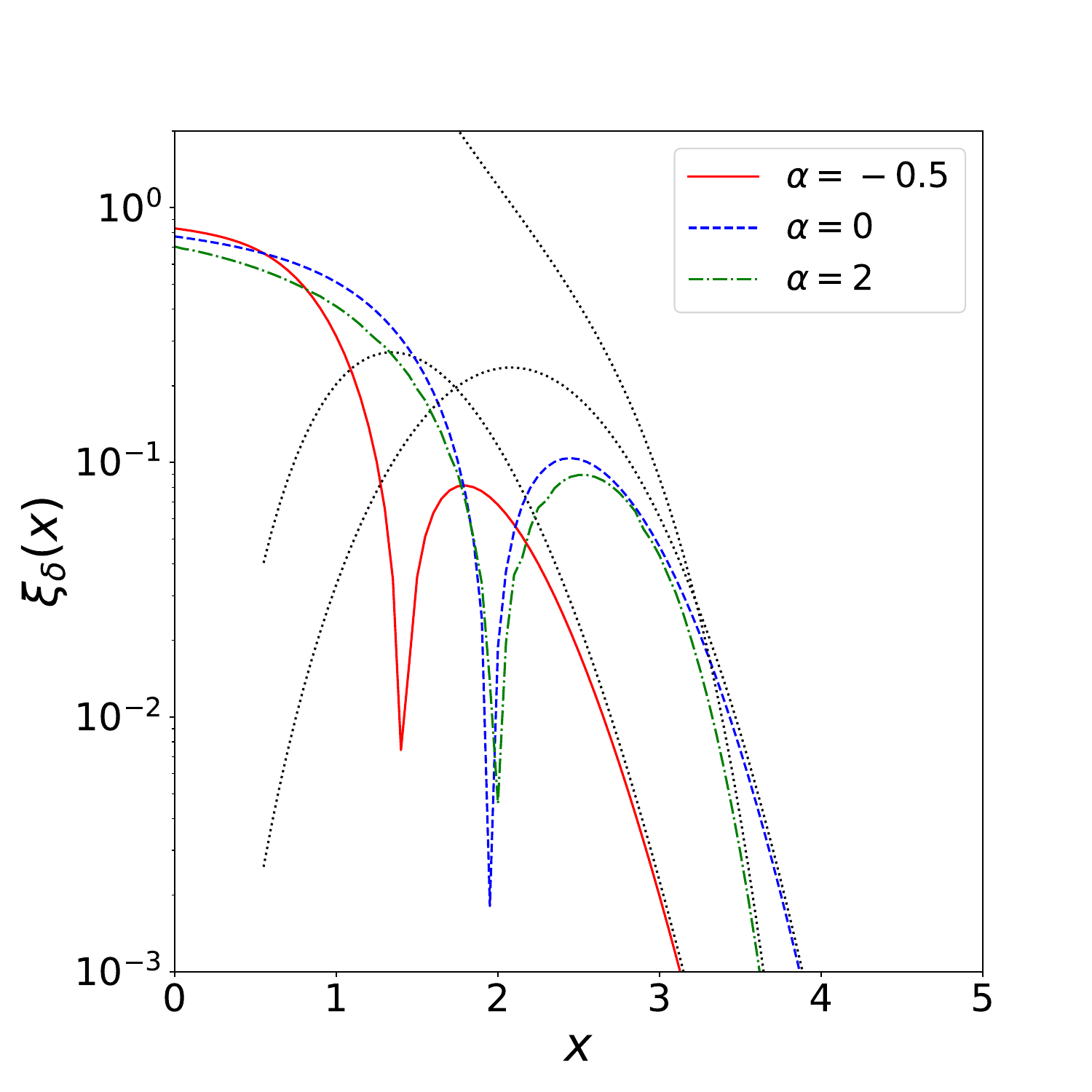}
\includegraphics[height=6.cm,width=0.32\textwidth]{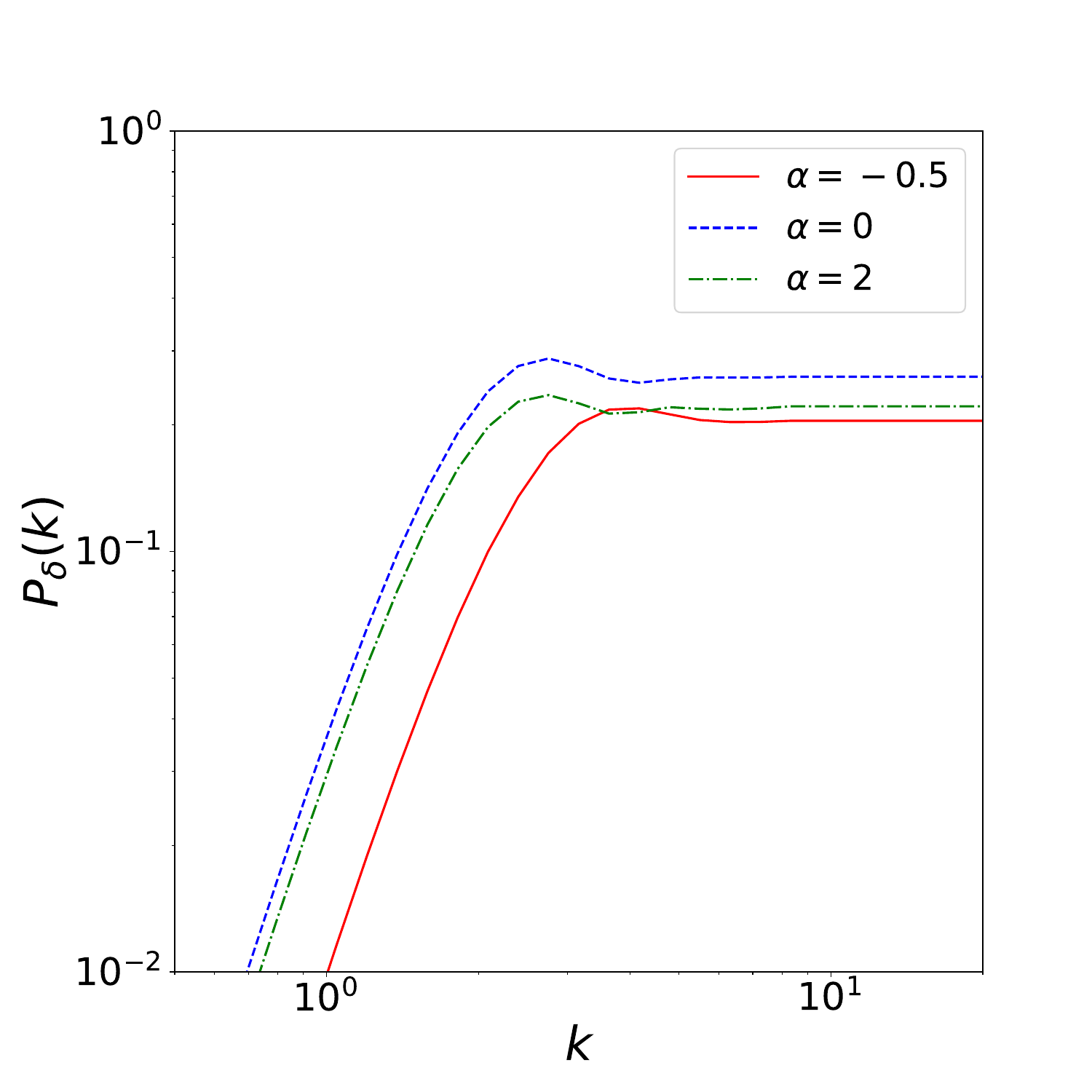}
\caption{
{\it Left and middle panels:} density correlation function $\xi_{\delta}^{\neq}(x)$ 
on linear and logarithmic scales for $x>0$.
{\it Right panel:} density power spectrum $P_{\delta}(k)$.
}
\label{fig:xi}
\end{figure}

The conservation of matter, encoded in the continuity equation, implies that the density field
$\rho(x)$ (normalized by the mean density of the system) and the density contrast $\delta=\rho-1$
are given by
\be
\rho(x) = \frac{dq}{dx} , \;\;\; \delta(x) = - \frac{dv}{dx} , \;\; \mbox{and} \;\; \langle \rho \rangle =1 , \;\;
\;\; \langle \delta \rangle =0 ,
\ee
in terms of the Eulerian map $q(x)$ and the velocity field $v(x)$, where we used $v(x)=x-q(x)$.
Defining the two-point density correlation function 
$\xi_\delta(x) = \langle \delta(x_1) \delta(x_1+x) \rangle$ and the associated
power spectrum $P_\delta(k)$, we obtain
\be
\xi_\delta(x) = - B_v''(x) , \;\; P_\delta(k) = k^2 E(k) ,
\ee
where $B_v(x)$ and $E(k)$ denote the velocity correlation function and the energy spectrum introduced
in Section~\ref{sec:energy-spectrum}.
Using the results in Eqs.(\ref{eq:Bv-deriv}) and (\ref{eq:Bv-asymp}), we find
\be
\xi_\delta(x) = \xi_0 \, \delta_D(x) + \xi_{\delta}^{\neq}(x) , \;\;\; \mbox{with for} 
\;\;\; x>0 : \; \xi_{\delta}^{\neq}(x) =  - \frac{1}{\alpha+1}  \frac{d^3}{dx^3} 
\left[ x {\cal R}_{\alpha+1}(x) \right] , 
\label{eq:xi-delta-deriv}
\ee
\be
\xi_0 = -\frac{4}{\alpha+1} {\cal R}'_{\alpha+1}(0) = \frac{16 \sqrt{2\pi} \Gamma(2\alpha+2) 
\Gamma\left( \frac{4\alpha+7}{2\alpha+3}\right)}{(\alpha+1)(\alpha+2) \Gamma(2\alpha+7/2)}
\Lambda_{\alpha}^{-(4\alpha+7)/(2\alpha+3)} > 0 ,
\label{eq:xi-0-def}
\ee
where the Dirac contribution $\xi_0 \, \delta_D(x)$ arises from the absolute-value term 
in Eq.(\ref{eq:Bv-asymp}) and accounts for the presence of shocks, while $\xi_{\delta}^{\neq}(x)$
is a finite and even function. 
These relations provide closed-form expressions and asymptotic behaviors of $\xi(x)$ and
$P_\delta(k)$. In particular,
\be
| x | \gg 1 : \xi_\delta(x) \sim x^{-2\alpha^2+9/2} \, e^{-\Lambda_\alpha 2^{-3\alpha-7/2} x^{2\alpha+3} } ,
\label{eq:xi-large-x}
\ee
\be
| k | \ll : \;\;\; P_{\delta}(k)\propto k^4, \;\;\;
| k | \gg 1 : \;\;\; P_{\delta}(k) \simeq \sqrt{\frac{2}{\pi}}  
\frac{8 \Gamma(2\alpha+2) \Gamma\left(\frac{4\alpha+7}{2\alpha+3}\right)} 
{(\alpha+1) (\alpha+2) \Gamma(2\alpha+7/2)} \Lambda_{\alpha}^{-(4 \alpha+7)/(2\alpha+3)}  .
\label{eq:Pk}
\ee

Figure~\ref{fig:xi} shows the density correlation function and the associated power spectrum.
The density correlation is negative at small separations, $x \ll 1$, reflecting the depletion of mass
in the Eulerian region surrounding shocks. It also becomes negative at large separation. 
Thus, although the initial conditions generated by the Poisson point process are completely uncorrelated, 
the nonlinear Burgers dynamics generate nontrivial correlations on large scales, which nevertheless 
decay as a stretched exponential.
Similar to the velocity power spectrum, the density spectrum exhibits only a weak dependence
on $\alpha$, as it follows the universal large-scale $k^4$ behavior and approaches a constant 
at high wavenumbers, consistent with Eq.(\ref{eq:Pk}).

\subsection{Lagrangian increment}
\label{sec:Lagrangian-increment}

\begin{figure}
\centering
\includegraphics[height=6.cm,width=0.48\textwidth]{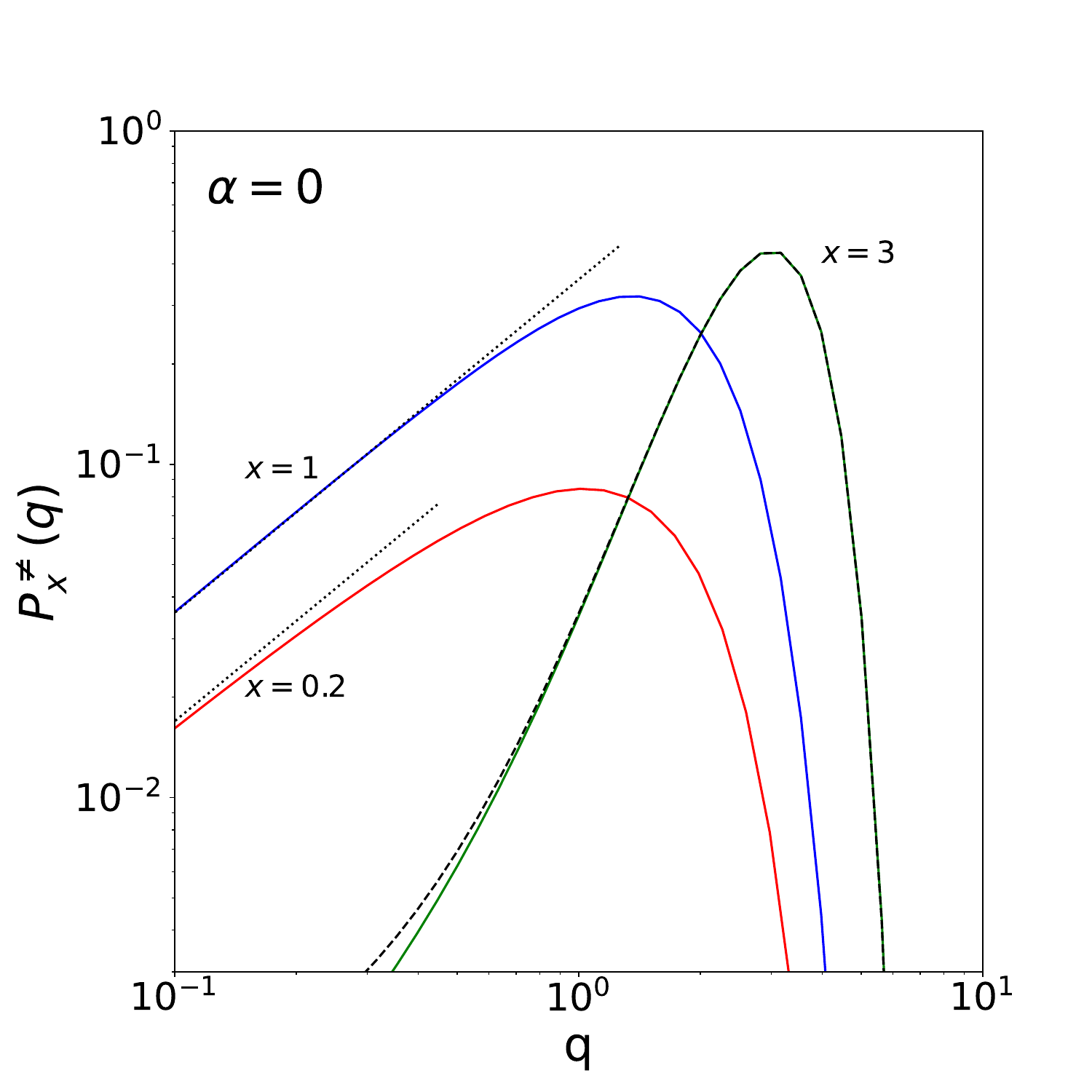}
\includegraphics[height=6.cm,width=0.48\textwidth]{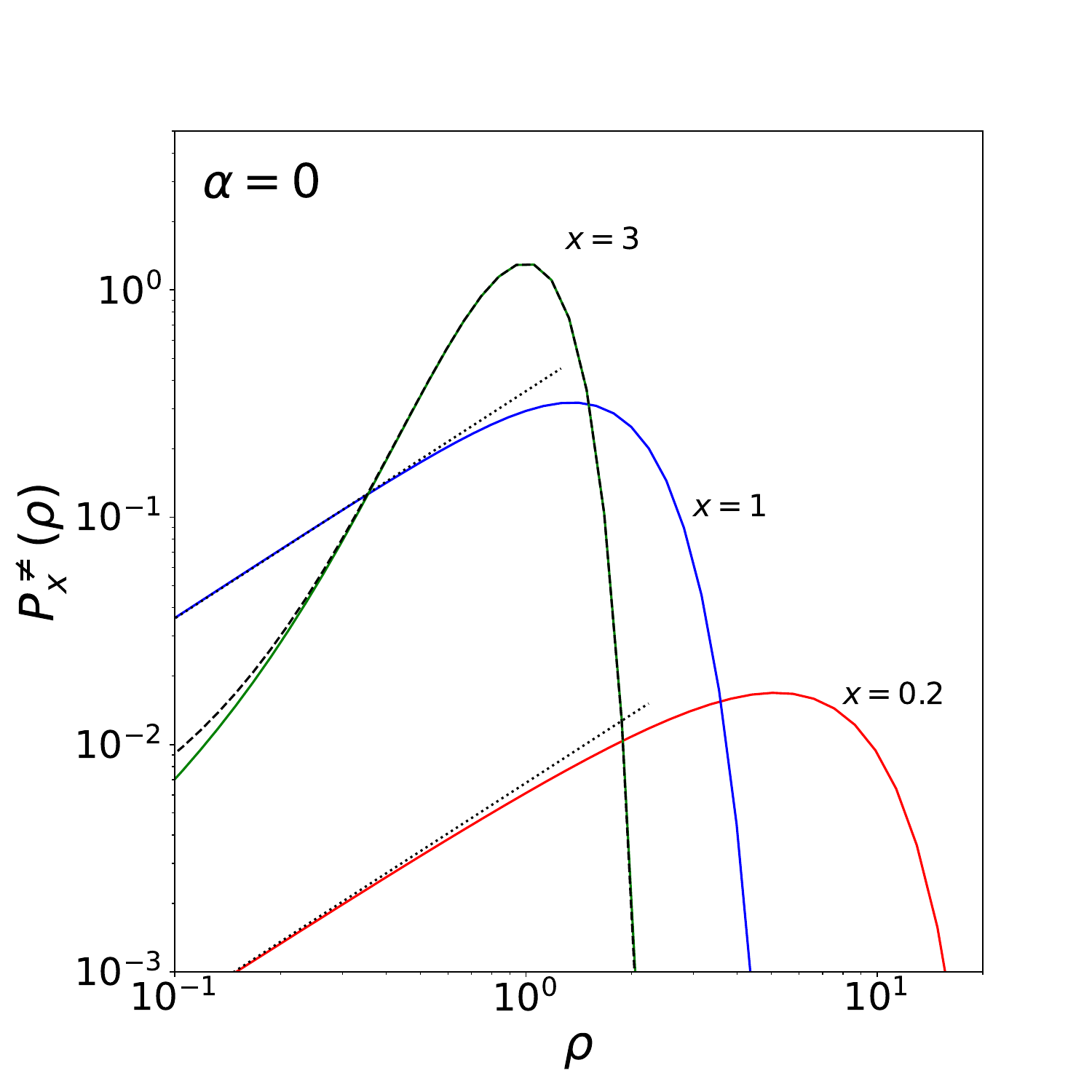}
\caption{
{\it Left panel:} probability distribution $P_x^{\neq}(q)$ for the case $\alpha=0$, for the three scales 
$x= 0.2, 1$, and $3$.
The black dotted lines are the small-$q$ linear asymptote (\ref{eq:P-x-q-asymp}) and the
black dashed line for $x = 3$ is the large-separation asymptote (\ref{eq:P_x_q-large-peak}).
{\it Right panel:} probability distribution $P_x^{\neq}(\rho)$ for the same cases.}
\label{fig:P_x_q}
\end{figure}

We now consider the distribution of the Lagrangian increment $q=q_2-q_1=q'_2-q'_1$ 
associated with the Eulerian interval $[x_1,x_2]$. It is defined by
\be
x = x_2 - x_1 \geq 0 , \;\; q = q_2 - q_1 \geq 0 : \;\;  P_x(q) = \int_{-\infty}^{\infty} 
dq'_1 dq'_2 \, P_x(q'_1,q'_2) \delta_D(q'_2-q'_1-q) .
\label{eq:P-x-q-inc-def}
\ee
From Eq.(\ref{eq:Pq1q2}), this distribution contains a Dirac contribution—corresponding to the
case $q_1=q_2=q_\star$—and a regular part,
\be
P_x(q) = P_{\rm void}(x) \, \delta_D(q) + P^{\neq}_x(q) ,
\label{eq:P-x-q-split}
\ee
with
\be
P^{\neq}_x(q) = x \int_{-\infty}^{\infty} dq'_\star \int_{\psi_{\min}(q'_\star)}^{\infty} 
\!\! d\psi_\star \, e^{- {\cal I}(\psi_\star,q'_\star)} \int_{q'_\star-q/2}^{q'_\star+q/2} 
d q' \, \psi_-(q'-q/2)^{\alpha} \, \psi_+(q'+q/2)^{\alpha}  .
\label{eq:P-neq-def}
\ee
Throughout the paper, we adopt the convention $\int_0^\infty dq \delta_D(q)=1$, i.e.,
the Dirac distribution contributes unit mass when integrated over $q\geq 0$.
The Dirac component is fully determined by the void probability studied in 
Section~\ref{sec:Void-probabilities}; therefore, in this section we focus on the regular part
$P^{\neq}_x(q)$.
The total weight of this regular contribution decreases linearly with $x$ at small separations,
since $P_{\rm void} \to 1$, and satisfies
\be
x \to 0 : \;\;  \int_0^{\infty} dq \, P^{\neq}_x(q) = 1 - P_{\rm void}(x) = N_{\rm void} x + \dots , 
\;\;\;\; x \to \infty : \;\;  \int_0^{\infty} dq \, P^{\neq}_x(q) \to 1 .
\label{eq:Px_q_x0}
\ee
Using $q_2'-q_1'= x + \frac{\partial \psi_-}{\partial q'_1} - \frac{\partial \psi_+}{\partial q'_2}$
and integration by parts, one can verify that the first moment obeys $\langle q \rangle_x = x$.

From (\ref{eq:P-neq-def}), we obtain the small-$q$ linear behavior at fixed $x$,
\be 
q \to 0 : \;\; P_x^{\neq}(q) = {\cal R}_{2\alpha}(x) \, x q  .
\label{eq:P-x-q-asymp}
\ee
For large $x$ and $q$ at fixed $|q-x|$, we obtain
\be
x \to \infty , \;\; q \to \infty , \;\; |q-x| \sim 1 : \;\;
P_x^{\neq}(q) \simeq f_{\infty}^{\neq}(|q-x|) \;\;\; \mbox{with} \;\;\;
f_{\infty}^{\neq}(q) = \int_{-\infty}^{\infty} dq' P_0(q'+q/2) P_0(q'-q/2) ,
\label{eq:P_x_q-large-peak}
\ee
where $P_0$ is the one-point distribution (\ref{eq:P_0-q}).
This is the large-separation limit $x \gg 1$, where the statistics at $x_1$ and $x_2$ become 
independent, as Eq.(\ref{eq:P_x_q-large-peak}) also reads
$P_x^{\neq}(q) \simeq \int_{-\infty}^{\infty} dq_1 dq_2 \delta_D(q_2-q_1-q) P_0(q_1-x_1) P_0(q_2-x_2)$.
For small separations $x$ at fixed $q$, we obtain
\be
x \to 0 : \; P_x^{\neq}(q) \simeq x \, n_{\rm shock}(q) 
\label{eq:P-x-q-small-x}
\ee
with
\be
n_{\rm shock}(q) = q \! \int_{q^2/8}^{\infty} \!\! dc \, e^{-\Lambda_\alpha c^{\alpha+3/2}} \!\!
\int_{q/2-\sqrt{2c}}^{-q/2+\sqrt{2c}} \!\!\! dq' [c \!-\! (q' \!-\! q/2)^2/2]^{\alpha} [c \!-\! (q' \!+\! q/2)^2/2]^{\alpha} .
\label{eq:nq-shocks}
\ee
We will show in Section~\ref{sec:shocks} that $n_{\rm shock}(q)$ is precisely the shock mass function. 
Thus, in the limit $x \to 0$, the distribution $P_x(q)$ is governed by the probability of finding either
zero shock (captured by $P_{\rm void}(x) \, \delta_D(q)$) or one shock (captured by $n_{\rm shock}(q)$).
This reflects the fact that all the matter is contained within discrete shocks.
The small-$x$ factorization (\ref{eq:P-x-q-small-x}) is consistent with the normalization
$\langle q \rangle_x = x$, using Eq.(\ref{eq:nshock-all-mass}) below, which states that all matter is
contained within shocks.

The left panel in Fig.~\ref{fig:P_x_q}  shows the regular part $P^{\neq}_{x}(q)$ for the case 
$\alpha=0$. 
At large separations $x \gg 1$, the regular component carries almost the full probability mass and the distribution becomes sharply peaked around its mean $\langle q \rangle = x$,
the approximation (\ref{eq:P_x_q-large-peak}) becoming increasingly accurate near the peak and 
over an expanding range.
Since the typical displacement of fluid elements is set by the scale $L(t)$ of Eq.(\ref{eq:L-t}),
on scales much larger than $L(t)$ the system remains nearly static, implying $x/q \to 1$.

On small scale $x \ll 1$, the total weight of $P^{\neq}_{x}(q)$ decreases linearly with $x$ 
as in (\ref{eq:Px_q_x0}), while the peak remains at $q \sim 1$.
This reflects the normalization of the full distribution to unity and $\langle q \rangle = x$.
The peak position is therefore independent of $x$ and instead set by the characteristic displacement 
scale $L(t)$, corresponding to $|x-q| \sim 1$ in the rescaled units (\ref{eq:re-scaling}).
Thus, as shown in Eq.(\ref{eq:P-x-q-small-x}), for $x \ll 1$ the regular part converges toward the fixed
shape $n_{\rm shock}(q)$, with an amplitude proportional to $x$.

Finally, the mean density $\rho$ within the Eulerian interval $[x_1,x_2]$ of length $x$ is
\be
\rho = \frac{q_2-q_1}{x_2-x_1} = \frac{q}{x} , \;\; \mbox{and} \;\; 
P_x(\rho) = P_{\rm void}(x) \delta_D(\rho) + P^{\neq}_x(\rho) \;\; \mbox{with} \;\;
P^{\neq}_x(\rho) = x P^{\neq}_x(q) , 
\ee
where $q=q_2-q_1$ is the Lagrangian increment as in Section~\ref{sec:Lagrangian-increment}.
The right panel of Fig.~\ref{fig:P_x_q}  shows the regular component $P^{\neq}_{x}(\rho)$ for the case 
$\alpha=0$. 
We again clearly see the linear decrease with $x$ of the total weight of the regular component 
$P^{\neq}_x(\rho)$, but the peak now shifts toward high density contrasts, $\rho \sim 1/x$.
This trend reflects the increasing inhomogeneity of the system on small Eulerian scales, arising from the
alternating sequence of shocks and voids.
At large separation, by contrast, the system appears progressively more homogeneous, and the distribution
becomes increasingly concentrated around the mean density.

\section{Higher-order distributions}
\label{sec:higher-order}

Higher-order distributions provide a more complete statistical description but are often
difficult to compute in nonlinear systems. 
As seen below in Eq.(\ref{eq:Pn-prod}), a key property of the Burgers dynamics for the class 
of initial conditions considered here is that these distributions exhibit a simple factorized 
structure, which significantly reduces their complexity.
Such explicit expressions and factorizations can only be derived for a few other cases,
see, for example \cite{Molchanov1995} for the Gaussian case and
\cite{Valageas2025} for Fr\'echet-type Poisson cases.
As for the two-point distribution (\ref{eq:Pq1c1q2c2}), the analysis simplifies when we consider
together the two variables $(q_i,c_i)$ associated with an Eulerian point $x_i$.
Owing to the uncorrelated nature of the underlying initial Poisson point process and to the ordering
$q_1 \leq q_2 \leq \cdots \leq q_n$ for $x_1 < x_2 < \cdots < x_n$,
the $n$-point distribution can be written as
\be
x_1 < x_2 < \cdots < x_n : \;\;\; P_{x_1 ,\cdots,x_n}(q_1,c_1; \cdots ;  q_n,c_n) 
= P_{x_1}(q_1,c_1) \prod_{i=2}^n P_{x_i,x_{i-1}}(q_i,c_i | q_{i-1},c_{i-1}) ,
\label{eq:Pn-prod}
\ee
where $P_{x_1}(q_1,c_1) = P_0(q_1-x_1,c_1)$ is the one-point distribution obtained in
Section~\ref{sec:one-point-Eulerian}, and $P_{x_i,x_{i-1}}(q_i,c_i | q_{i-1},c_{i-1})$ is the conditional
probability
\bea
P_{x_2,x_1}(q_2,c_2 | q_1,c_1) & = & \left[ \delta_D(q_2-q_1) \delta_D(c_2-c_{12}) + \theta(q_2>q_{12})
\theta(c_2<c_{12}) \lambda( c_2 - (q_2-x_2)^2/2) \right] \nonumber \\
&& \times e^{ - \int_{q_{12}}^{\infty} dq \int_{c_1-(q-x_1)^2/2}^{c_2-(q-x_2)^2/2} d\psi \, \lambda(\psi) } .
\label{eq:P21-conditional}
\eea
One may verify that this expression reduces to Eq.(\ref{eq:Pq1c1q2c2}) for the two-point distribution.
Here $c_{12}$ denotes the height of the parabola ${\cal P}_{x_2,c_2}$ that
intersects the previous parabola ${\cal P}_{x_1,c_1}$ at the position $q_1$.
For arbitrary height $c_2$, we changed notation from $q_\star$ to $q_{12}$ for the intersection of both parabolas,
\be
c_{12} = c_1 + \frac{ (q_1-x_2)^2 - (q_1-x_1)^2}{2} , \;\;\;
q_{12}(c_2) = \frac{x_1+x_2}{2} + \frac{c_1-c_2}{x_2-x_1} .
\ee
The normalization is easily checked, $\int dc_2 dq_2 P_{x_2,x_1}(q_2,c_2 | q_1,c_1) =1$.

The factorization (\ref{eq:Pn-prod}) holds because, for example in the three-point case, one can
verify that $q_{13} \geq q_{12}$.
Thus, the intersection of ${\cal P}_1$ and ${\cal P}_3$ is irrelevant:
it always occurs in a region where ${\cal P}_2$ already dominates ${\cal P}_1$, i.e.,
${\cal P}_2 \geq {\cal P}_1$.
Consequently, the parabolic arcs follow the same ordering $({\cal P}_1, \cdots , {\cal P}_n)$
as the Eulerian points $(x_1,x_2,\cdots,x_n)$ over the domains 
$-\infty < q_{12} \leq q_{23} \leq \cdots \leq q_{n-1,n} < \infty$.
Therefore, when adding a new point $x_n$ in the $n$-point distribution (\ref{eq:Pn-prod}),
it is sufficient to consider only the preceding parabola ${\cal P}_{x_{n-1},c_{n-1}}$ and its contact
point $q_{n-1}$.
As in the two-point case (\ref{eq:Pq1c1q2c2}), one must distinguish whether the new contact 
point $q_n$ coincide with $q_{n-1}$—the first term in Eq.(\ref{eq:P21-conditional}),
which prevents assigning a double Poisson weight to the same contact point—or whether it lies to the right,
corresponding to the second term.
Finally, the exponential factor accounts for the requirement that the region in the $(q,\psi)$ plane 
above the preceding parabola ${\cal P}_{n-1}$ and below the new parabola
${\cal P}_{n}$ contains no Poisson points.

The factorization (\ref{eq:Pn-prod}) implies that all multi-point statistics can be
reconstructed from a finite set of basic building blocks.
In practice, it means that once the two-point conditional distributions are known, higher-order 
correlations carry no independent information beyond combinatorial structure.
Such hierarchical structures are also encountered in other areas 
but they are rarely derived from first principles in a fully nonlinear stochastic system.
For instance, in cosmology hierarchical models \cite{Peebles-1980,Balian-1989,Bernardeau1992}
are introduced as a phenomenological ansatz to describe the large-scale distribution of matter. 
Within some approximations, there have been attempts \cite{Fry_1984} to relate the associated parameters to the BBGKY equations that govern the hierarchy of 
correlation functions, but these hierarchical models are not an exact solutions of the dynamics.
In the Burgers dynamics that we study here, they are exact results that emerge naturally
from the geometry of first-contact parabolas and the Poisson nature of the initial potential.

\section{Lagrangian distributions or particle displacements}
\label{sec:Lagrangian}

In this Section, we finally turn to Lagrangian statistics, which track individual fluid elements
rather than fixed Eulerian positions.
This viewpoint is particularly suitable for studying shocks, because shocks are precisely the
locations where many Lagrangian trajectories converge and mass is concentrated.
We focus on the distribution of Eulerian
positions $x$ for particles with prescribed Lagrangian coordinates $q$.
Equivalently, this corresponds to the statistics of the displacements $x-q$, where each particle is 
labeled by its initial positions $q$.
This displacement field directly encodes how far particles have been transported by the flow,
and its distribution reveals the relative importance of small, frequent displacements versus rare, 
large excursions.

\subsection{One-point Lagrangian distribution}
\label{sec:one-point-Lagrangian}

Because particles do not cross, the Lagrangian probability $P_q(\geq x)$ that the particle labeled by
$q$ lies to the right of position $x$ is equal to the Eulerian probability $P_x(\leq q)$
that the Lagrangian coordinate $q(x)$ of the particle located at $x$ is less than or equal to $q$.
Differentiating this relation with respect to $x$, and using the identity
$P_x(q)=P_0(q-x)$ derived in Section~\ref{sec:one-point-Eulerian}, yields
\be
P_q(\geq x) = P_x(\leq q)  \;\;\; \mbox{and} \;\;\; P_q(x) = P_0(x-q) = P_x(q) .
\ee
Thus, the one-point Lagrangian and Eulerian distributions are identical, and all properties
of $P_q(x)$ follow directly from the results obtained in Section~\ref{sec:one-point-Eulerian}.

\subsection{Two-point Lagrangian distribution}
\label{sec:two-point-Lagrangian}

As in the case of one-point statistics, the two-point Lagrangian and Eulerian probabilities are related
by
\be
P_{q_1,q_2}(\geq x_1,\leq x_2) = P_{x_1,x_2}(\leq q_1,\geq q_2) , \;\;\;
P_{q_1,q_2}(x_1,x_2) = - \frac{\partial^2}{\partial x_1 \partial x_2} \int_{-\infty}^{q_1} dq_1''
\int_{q_2}^{\infty} dq_2'' \, P_{x_1,x_2}(q_1'',q_2'') .
\label{eq:P-twopoint-Lag}
\ee
Using statistical homogeneity, which implies $P_{x_1,x_2}(q_1,q_2) = P_x(q_1-\bar x,q_2-\bar x)$ 
as in Eq.(\ref{eq:Pq1q2}), and changing variables from $(x_1,x_2)$ to
$(x,\bar x)$, we obtain
\be
P_{q_1,q_2}(x_1,x_2) = \left( \frac{\partial^2}{\partial x^2} - \frac{1}{4} \frac{\partial^2}{\partial \bar x^2}
\right) \int_{-\infty}^0 dq_1'  \int_{q}^{\infty} dq_2' \, P_x(q_1'+q_1-\bar x,q_2'+q_1-\bar x) ,
\ee
where $q=q_2-q_1$.
As in Section~\ref{sec:Lagrangian-increment}, we focus on the probability
distribution of the Eulerian increment $x$,
\be
P_q(x) = \int dx_1 dx_2 \, P_{q_1,q_2}(x_1,x_2) \, \delta_D(x_2-x_1-x) .
\ee
Changing variables again from $(x_1,x_2)$ to $(x,\bar x)$ and making use of (\ref{eq:P-x-q-inc-def}),
we obtain the relation between the distributions of Lagrangian and Eulerian increments,
\be
q > 0 : \;\;\; P_q(x) = P_{\rm shock}(q) \, \delta_D(x) + P_q^{\neq}(x) \;\;\; \mbox{with} \;\;\;
P_q^{\neq}(x) = \frac{\partial^2}{\partial x^2} \int_q^{\infty} dq' \, P_x^{\neq}(q') (q'-q) ,
\label{eq:P-q-x-def}
\ee
where the Dirac term in Eq.(\ref{eq:P-x-q-split}) does not
contribute for $q>0$, and we have explicitly added the term $P_{\rm shock}(q) \delta_D(x)$, 
which corresponds to the probability that the entire interval $[q_1,q_2]$ lies within a single shock.
The latter contribution is not included in the regular term $P_q^{\neq}(x)$.
Writing the integral $q$ as $\int_q^{\infty}=\int_0^{\infty}-\int_0^q$ and using 
Eqs.(\ref{eq:P-x-q-split}) together with $\langle q \rangle_x=x$, we can rewrite
\be
P_q^{\neq}(x) = q \, n_{\rm void}(x) + 
\frac{\partial^2}{\partial x^2} \int_0^q dq' \, P_x^{\neq}(q') (q-q') , \;\; \mbox{whence} \;\;
\langle x \rangle_q = q ,
\label{eq:P-q-x-void}
\ee
where the equality for the mean follows from integrations by parts and the result $P_{\rm void}(0)=1$
from Eq.(\ref{eq:Pvoid-x-0-large-x}).
This confirms that the system exhibits no net expansion or contraction: particle displacements occur 
on the scale $L(t)$ introduced in Eq.(\ref{eq:L-t}), and therefore $x/q \to 1$ for $q\to\infty$.
Consequently, $\langle x \rangle_q=q$ for any Lagrangian interval.

For small Lagrangian separations at fixed $x$ we find
\be
q \to 0 : \;\;\; P_q^{\neq}(x) \simeq q \, n_{\rm void}(x) .
\label{eq:Pq-x-small-q}
\ee
This factorized form is analogous to Eq.(\ref{eq:P-x-q-small-x}) for the Eulerian
distribution $P_x^{\neq}(q)$ at small $x$.
It shows that, for small Lagrangian mass intervals $q \to 0$, the Eulerian distance distribution 
is determined, to order $q$, by the probability of being fully contained within
a shock (contributing the term $P_{\rm shock}(q) \, \delta_D(x)$) or by containing a single
void (described by the void multiplicity function $n_{\rm void}(x)$).
This reflects the discreteness of voids in Lagrangian space and the fact that they occupy all of the
Eulerian volume.
The normalization $\langle x \rangle_q = q$, together with Eq.(\ref{eq:nvoid-power-law}), is fully consistent 
with the small-$q$ behavior in Eq.(\ref{eq:Pq-x-small-q}).

At large scales, the Eulerian distribution takes the form $P_x^{\neq}(q) \simeq f_\infty^{\neq}(|q-x|)$,
as given by Eq.(\ref{eq:P_x_q-large-peak}).
Substituting this into Eq.(\ref{eq:P-q-x-def}) yields
\be
q \to \infty , \;\; x \to \infty , \;\; |x-q| \sim 1 : \;\;
P_q^{\neq}(x) \simeq f_{\infty}^{\neq}(|x-q|) \simeq P_x^{\neq}(q) .
\label{eq:P_q_x-large-peak}
\ee
As expected, the distribution is sharply peaked at the mean $\langle x \rangle_q=q$,
with a width set by the characteristic displacement scale of the particles.

The left panel of Fig.~\ref{fig:nshock} shows the distribution $P_q^{\neq}(x)$.
As in the Eulerian case displayed in Fig.~\ref{fig:P_x_q}, the weight of this regular component
decreases linearly with $q$ on small scales, in agreement with Eq.(\ref{eq:Pq-x-small-q}), 
while its characteristic width—visible through the sharp cutoff—remains of order unity and is set
by the typical void size.
The main difference relative to the Eulerian distribution is that $P_q^{\neq}(x)$ does not decrease 
linearly with $x$ at small separations; instead it approaches a finite constant as $x \to 0$.
At large separations, the distribution becomes again increasingly sharply peaked around its mean
$\langle x \rangle_q=q$, and Eulerian and Lagrangian statistics converge, with 
$P_q^{\neq}(x) \simeq P_x^{\neq}(q)$.

\subsection{Multiplicity function of shocks}
\label{sec:shocks}

\begin{figure}
\centering
\includegraphics[height=6cm,width=0.33\textwidth]{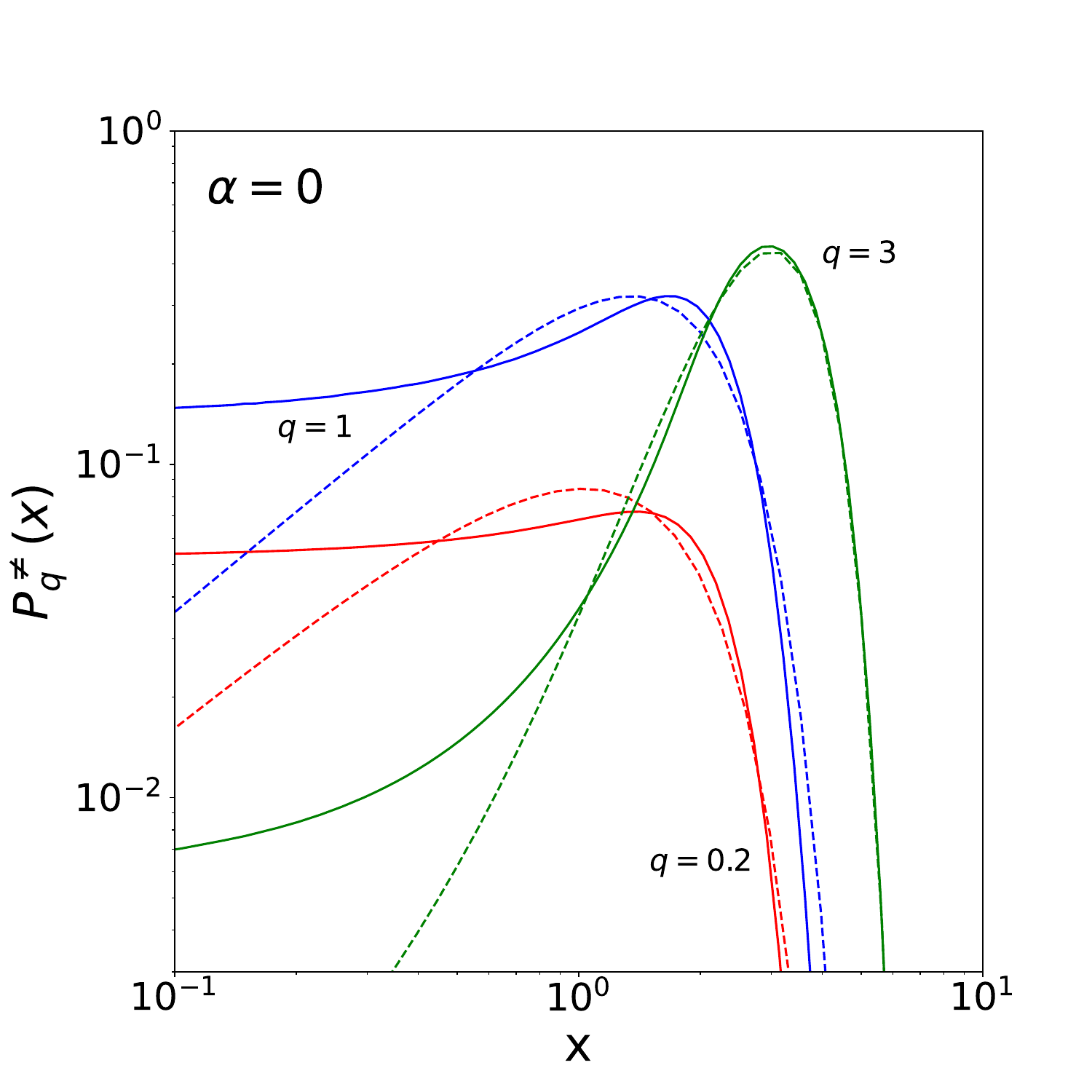}\includegraphics[height=6cm,width=0.33\textwidth]{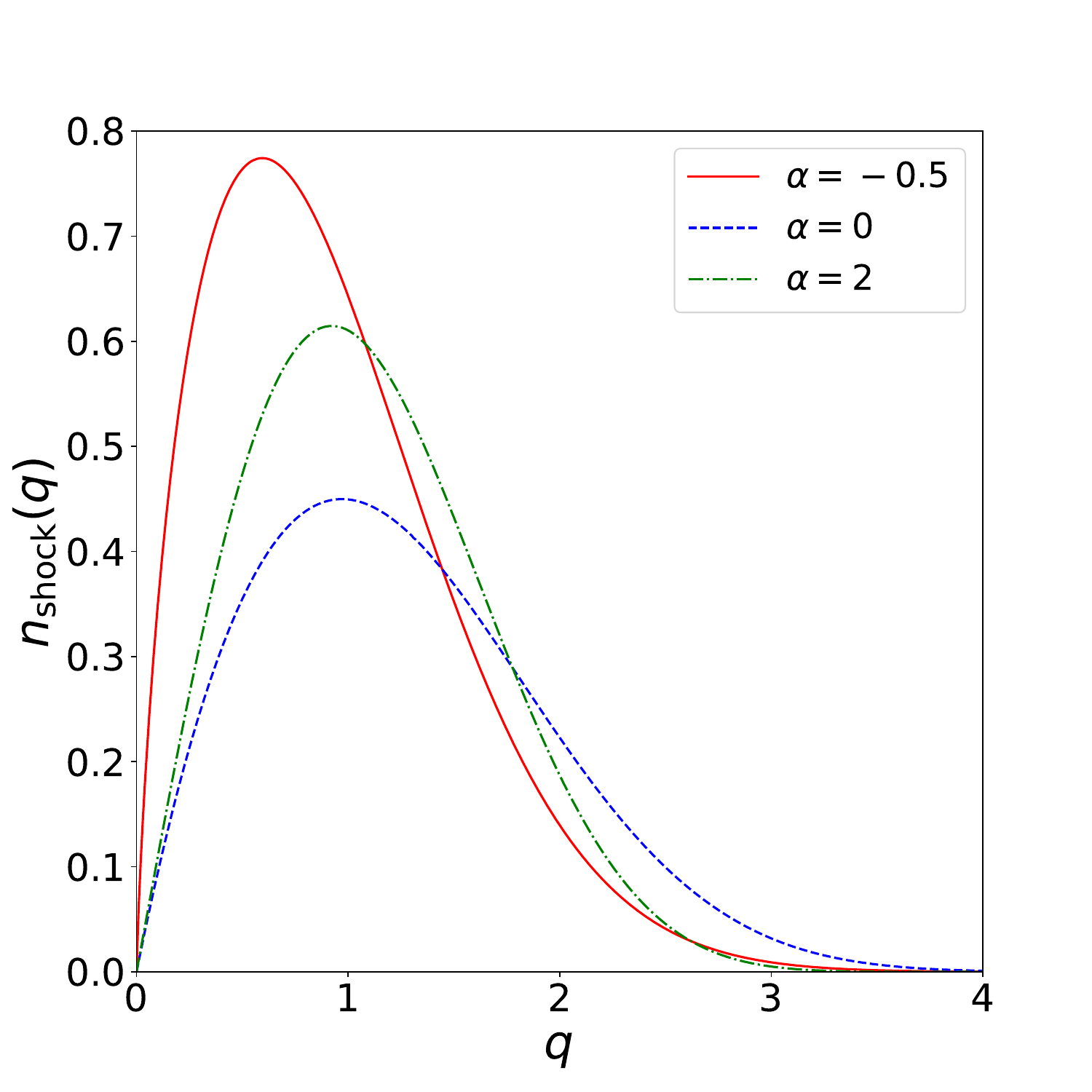}
\includegraphics[height=6cm,width=0.33\textwidth]{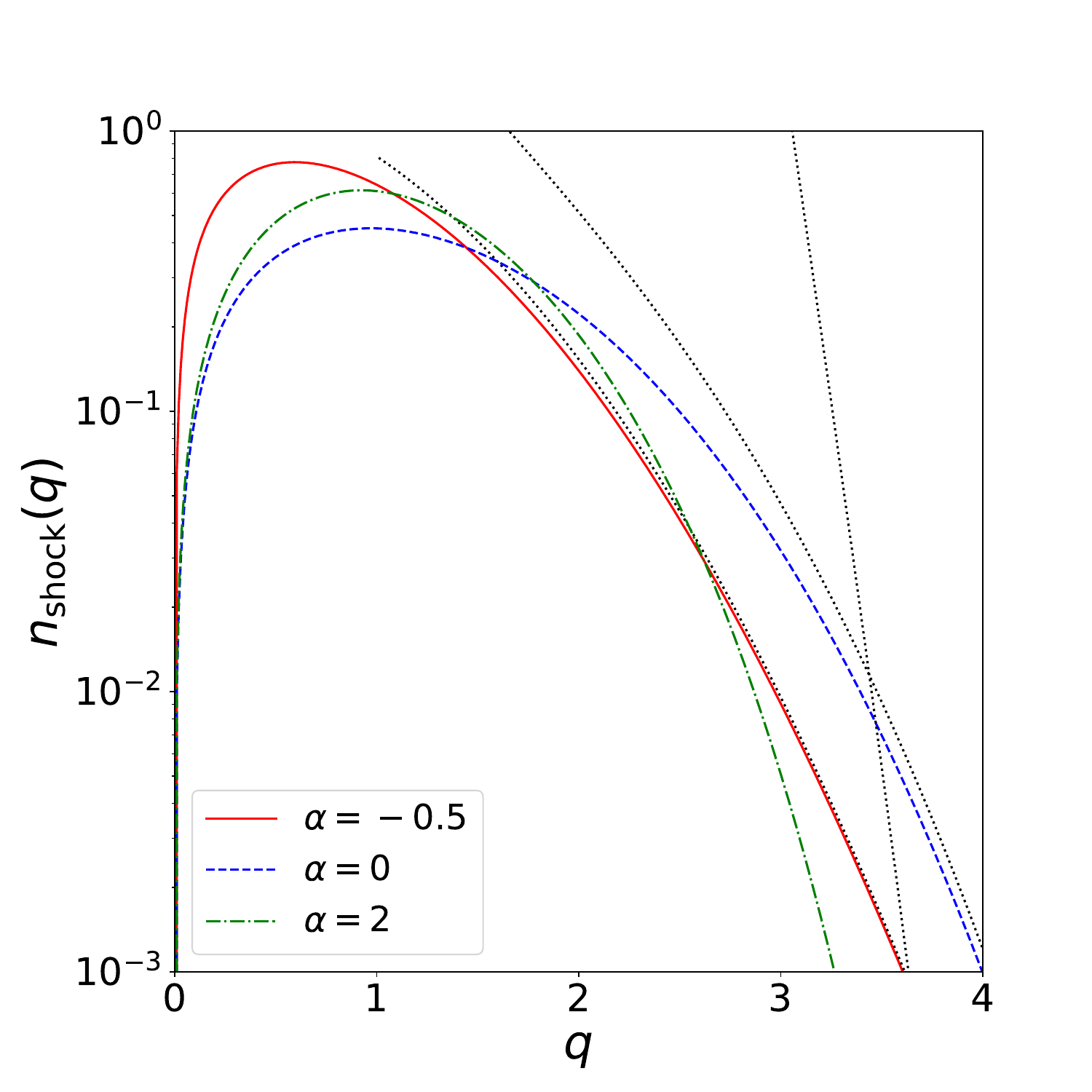}
\caption{
{\it Left panel:} probability distribution $P_q^{\neq}(x)$ of the Eulerian increment $x$,
for the case $\alpha=0$ and the three scales $q= 0.2, 1$, and $3$, as in Fig.~\ref{fig:P_x_q}.
{\it Middle and right panels:} shock multiplicity function from Eq.(\ref{eq:nq-shocks}) for the cases 
$\alpha=-0.5, 0$, and $2$.
}
\label{fig:nshock}
\end{figure}

To quantify the mass content of shocks, we introduce the shock multiplicity function
$n_{\rm shock}(q)$, which counts the number of shocks per unit length as a function of their mass.
This is the Lagrangian analogue of the void multiplicity function and plays a role similar to that
of a cluster or halo mass function in other aggregation problems.
First, we note that the probability $P_{\rm shock}(q)$ that a Lagrangian interval of size $q$
is entirely contained within a single shock follows from the normalization of the full distribution 
$P_q(x)$ in Eq.(\ref{eq:P-q-x-def}):
\be
P_{\rm shock}(q) = 1 - \int_0^{\infty} dx P_q^{\neq}(x) = 1 - N_{\rm void} \, q 
+ \left. \frac{\partial}{\partial x} \right|_{x=0} \int_0^q dq' \, P_x^{\neq}(q') (q-q') , 
\ee
where we have used Eq.(\ref{eq:P-q-x-void}).
Then, the quantity $P_{\rm shock}(q)$ is also related to the shock multiplicity function 
$n_{\rm shock}(q) dq$, defined as the number of shocks per unit length with Lagrangian mass in the
interval $[q,q+dq]$, through
\be
P_{\rm shock}(q) = \int_q^{\infty} dq' n_{\rm shock}(q') \, (q'-q) , \;\;\; \mbox{whence} \;\;
n_{\rm shock}(q) = \frac{d^2 P_{\rm shock}}{dq^2} = \left. \frac{\partial}{\partial x} \right|_{x=0} P_x^{\neq}(q) .
\ee
This immediately yields
\be
\int_0^{\infty} dq \, n_{\rm shock}(q) \, q = P_{\rm shock}(0) = 1 ,
\label{eq:nshock-all-mass}
\ee
which expresses the fact that all matter is bound within shocks.
Using the small-$x$ behavior of $P_x^{\neq}(q)$ from Eq.(\ref{eq:P-x-q-small-x}),
one recovers the expression for $n_{\rm shock}(q)$ given in Eq.(\ref{eq:nq-shocks}).
This provides, in particular, the small- and large-mass asymptotic falloffs
\be
q \to 0 : \;\; n_{\rm shock}(q) \simeq {\cal R}_{2\alpha}(0) \, q , \;\; \mbox{and for}  \;\;
q \to \infty: \;\; n_{\rm shock}(q) \sim q^{-2(\alpha+1) (2\alpha+1)} \,
e^{-\Lambda_\alpha 2^{-3\alpha-9/2} q^{2\alpha+3}} .
\label{eq:shock-asymp}
\ee
These asymptotic regimes show how the Poisson–Weibull initial conditions control the relative 
abundance of small versus massive shocks.
In particular, the sensitivity of the high-mass tail to $\alpha$ parallels the behavior of large voids 
and provides another manifestation of how the tail of the initial potential distribution shapes extreme 
events in the evolved flow.
The middle and right panels in Fig.~\ref{fig:nshock} show the shock multiplicity functions.
We recover the asymptotic regimes (\ref{eq:shock-asymp}).
Together with the Eulerian results, the Lagrangian statistics presented in this section complete 
our description of how mass is transported, compressed into shocks, and separated by voids in Burgers 
dynamics with Weibull-class initial conditions.
They offer a compact, analytically controlled summary of the coarsening process and the emergence of a 
shock-dominated medium from a simple Poisson ensemble of initial fluctuations.

In particular, an appealing interpretation of these results emerges when one views shocks as the 
macroscopic counterparts of high-density clusters or jams in exclusion or aggregation processes
\cite{Frachebourg2000,Valageas2009}. 
In many one-dimensional lattice gases, microscopic shocks and jams coarsen and merge under the 
hydrodynamic evolution, and the resulting distribution of cluster sizes controls the statistics of 
transport and current fluctuations.
The shock multiplicity function derived here thus provides an analytically solvable analogue of a
jam-size distribution in a noise-free Burgers limit, in which the tail of the initial potential 
distribution (encoded in $\alpha$) selects how likely large macroscopic clusters are.
Such solvable examples may inform coarse-grained descriptions of exclusion processes in regimes where 
ballistic motion dominates and stochasticity appears primarily through the initial condition.

\section{Conclusion}
\label{sec:conclusion}

In this paper, we have presented a detailed analytical characterization of the statistical properties
of one-dimensional Burgers dynamics for a class of initial conditions generated by a Poisson point 
process with a power-law intensity of exponent $\alpha > -1$. 
This family leads to Weibull-type statistics, featuring stretched-exponential tails,
whose exponents vary continuously with $\alpha$ and span the range from 1 to $\infty$.
This demonstrates how changing the tail of the initial potential distribution directly translates 
into different degrees of intermittency in the evolved fields, while preserving self-similarity.
This class is therefore complementary to the case of Fr\'echet-type statistics
with power-law tails, investigated in the companion paper \cite{Valageas2025}, which arises when
the Poisson intensity follows a power-law with exponent $\alpha < -3/2$.
The two cases together provide a unified framework covering the full range of admissible exponents 
and reveal how the sign of the intensity exponent leads to qualitatively distinct statistical regimes 
in Burgers turbulence.
Taken together, these results provide a unified framework that clarifies how the statistical 
properties of Burgers turbulence depend on the tail behavior of the initial conditions. 
In particular, they highlight how rare events—encoded in the initial distribution—control
the asymptotic properties of the evolved system.
Beyond the explicit calculations, a central outcome of this work is to show that a highly nontrivial nonlinear system—featuring shocks, intermittency, and strong non-Gaussianity—can nevertheless admit a fully tractable statistical description for a suitable class of initial conditions.

Exploiting the geometrical interpretation of the solution in terms of parabolic first-contact points, 
we obtained exact expressions for a wide class of Eulerian and Lagrangian observables, 
including one-point and multi-point distributions, shock and void multiplicity functions, and the full probability 
density of Lagrangian and Eulerian increments. 
The formalism applies to arbitrary exponent $\alpha > -1$ and it includes, as a limiting case with 
a proper rescaling, the traditional case of Gaussian initial conditions without large-scale power
\cite{Kida1979}.

We established closed-form expressions for the shock mass function and void-size distribution, 
together with their large-mass and large-size asymptotic behaviors. The derived relations explicitly
show how these quantities control the small-scale structure of Eulerian and Lagrangian increments, 
leading to simple factorized forms in the limits of small intervals. On large scales, we demonstrated
that both Eulerian and Lagrangian increment statistics converge to peaked distributions governed
by the one-point distribution $P_0(q)$, with widths determined by the characteristic displacement 
scale $L(t)$.
These statistics also quantify how the line is partitioned into empty regions and shocks of 
various masses, and they reveal that voids fill essentially all Eulerian space, whereas matter is 
concentrated into a discrete set of shocks whose statistics are explicitly computable.
The resulting picture is that of a self-similar, highly intermittent medium with short-tailed fluctuations, in sharp contrast with the heavy-tailed Fréchet case previously studied for
$\alpha < -3/2$.

In addition, the paper presents an exact factorization of the full $n$-point distributions into a sequence of 
two-point conditional probabilities. This property follows from the ordered structure of the parabolic 
envelopes and substantially simplifies the statistical description of the process. 
This factorization property is particularly significant, as it suggests that the apparent 
complexity of higher-order statistics can, in some cases, be reduced to simpler building blocks.
Such structures may prove useful in other contexts involving nonlinear stochastic dynamics.

Beyond their intrinsic mathematical interest, these results offer a controlled framework
to explore how the tail behavior of initial fluctuations affects the formation and statistics
of shocks, voids, and correlations in Burgers-like systems.
They may serve as benchmarks for numerical studies of turbulence and for simplified models
of structure formation in cosmology, where variants of the adhesion model are used to describe
the emergence of the cosmic web.
It would be interesting in future work to extend this analysis to higher-dimensional geometries,
to other classes of point processes, or to settings where small-scale forcing and finite viscosity 
compete with the self-similar dynamics studied here.

The analysis presented in this work has been carried out entirely within the deterministic 
framework of inviscid Burgers dynamics with Poisson–Weibull initial conditions.
Nevertheless, the underlying mechanism—random initial data transported by noise-free Euler 
hydrodynamics—also appears in a broad range of stochastic transport models.
In particular, exclusion processes and related one-dimensional lattice gases often admit hydrodynamic 
limits in which the macroscopic densities obey Euler-type conservation laws, so that 
ballistic transport is governed by the same nonlinear advective term as in our setting
\cite{Spohn-1991,Doyon-2025}.
Whereas macroscopic fluctuation theory (MFT) considers driven diffusive systems 
\cite{Bertini-2015},
ballistic macroscopic fluctuation theory (BMFT) focuses on ballistic processes 
\cite{Doyon2023,kethepalli2025,Yoshimura-2025}, with only  subleading diffusive effects.
In such systems, initial fluctuations are transported by deterministic Euler dynamics,
while  microscopic randomness is confined to the initial state and to diffusive or weakly noisy 
corrections.
The deterministic Burgers equation that we consider in this paper is a simple example
of hydrodynamical processes governed by random initial conditions.
As in aggregation models, if we follow the trajectories of shocks as discrete particles
they obey purely ballistic motion as in the fully inviscid Burgers equation.
However, the infinitesimal viscosity plays a critical role at collisions and leads to
the irreversible merging of shocks.
Therefore, the systems that we consider in this paper are at the boundary between
MFT and BMFT, as both advective and diffusive terms play key roles.
Thus, our framework provides an analytically tractable example in which the statistics of shocks, 
voids, and currents can be computed explicitly under a competition between
noise-free advection and irreversible aggregation due to a nonzero viscosity.
It would be interesting to see whether the present analysis, and more generally the
Burgers equation with infinitesimal viscosity, could serve as a basis to describe systems
at the boundary between MFT and BMFT. This is left for future works.



 



\bibliography{ref}

\end{document}